\documentclass[prb,twocolumn,altaffilletter,longbibliography,numerical,flushbottom,secnumarabic,superscriptaddress,floatfix,nobibnotes]{revtex4-2}

\usepackage{graphicx}
\usepackage{braket}
\usepackage{amsmath, amssymb}
\usepackage{booktabs}  
\usepackage{comment}
\usepackage[normalem]{ulem}
\usepackage{cancel}

\usepackage[breaklinks, pdftex, hyperfootnotes=true, pdfpagelabels, bookmarks, pageanchor]{hyperref}
\hypersetup{%
	colorlinks=true, linktocpage=true, pdfstartpage=1, pdfstartview=FitH, pdfborder={0 0 0},%
	breaklinks=true, pdfpagemode=UseNone, pageanchor=true, pdfpagemode=UseOutlines,%
	plainpages=false, bookmarksnumbered, bookmarksopen=true, bookmarksopenlevel=1,%
	hypertexnames=true, pdfhighlight=/O,
	urlcolor=blue, linkcolor=blue, citecolor=blue,
	}
\usepackage{orcidlink}

\newcommand\identity{1\kern-0.25em\text{l}}

\graphicspath{{figs/}}
 
\usepackage{soul}

\newcommand{\lt}[1]{\textcolor{magenta}{ #1}}

\usepackage[colorinlistoftodos]{todonotes}
\presetkeys{todonotes}{inline}{}

\newcommand{\tiltedparallel}{\mathbin{\!/\mkern-5mu/\!}}

\begin{document}
\title{Emergence of Bogoliubov Fermi Surfaces in hybrid Al/InAs heterostructures}
\newcommand{\affA}{\affiliation{Departamento de S\'olidos Cu\'anticos y Sistemas Desordenados, Centro At\'omico Bariloche, Instituto de Nanociencia y Nanotecnolog\'ia CONICET-CNEA and Instituto Balseiro (8400), San Carlos de Bariloche, Argentina.}}
\newcommand{\affB}{\affiliation{Grupo de Circuitos Cuánticos Bariloche, Div. Dispositivos y Sensores, Centro Atómico Bariloche-CNEA, Instituto Balseiro and CONICET, (8400) San Carlos de Bariloche, Argentina.}}

\newcommand{\affD}{\affiliation{Institut f\"ur Experimentelle und Angewandte Physik, University of Regensburg, 93040 Regensburg, Germany.}}
\newcommand{\affE}{\affiliation{Laboratoire de Physique des Solides (CNRS UMR 8502), Bâtiment 510, Université Paris-Sud, Orsay, 91405, France.}}

\newcommand{\affMa}{\affiliation{Microsoft Quantum Purdue, Purdue University, West Lafayette, IN, USA.}}
\newcommand{\affMb}{\affiliation{Elmore Family School of Electrical and Computer Engineering, Purdue University, West Lafayette, IN 47907 USA}}
\newcommand{\affMc}{\affiliation{Department of Physics and Astronomy, Purdue University, West Lafayette, IN, USA}}
\newcommand{\affMd}{\affiliation{School of Materials Engineering, Purdue University, West Lafayette, IN, USA}}
\newcommand{\affMe}{\affiliation{Purdue Quantum Science and Engineering Institute, Purdue University, West Lafayette, IN 47907 USA}}

\author{S. Feyrer}
\affD
\author{V. Dimic}
\affD
\author{I. Lobato}
\affD
\author{A. Kirchner}
\affD
\author{P. Drexler}
\affD
\author{L. Rupp}
\affD
\author{D. Bougeard}
\affD
\author{T. Lindemann}
\affMc
\author{S. Gronin} 
\affMa
\author{ G. Gardner} 
\affMa
\author{M. J. Manfra} 
\affMc \affMa \affMb \affMd \affMe
\author{G. F. R. Ruiz}
\affA
\author{C. A. Balseiro}
\affA
\author{L. Arrachea}
\affA
\author{M. Aprili}
\affE
\author{N. Paradiso}
\affD
\author{C. Strunk}
\affD
\author{L. Tosi}
\affB \affD
\date{\today}

\begin{abstract}
We investigate the microwave electrodynamics of a proximitized two-dimensional electron gas in hybrid superconductor/semiconductor heterostructures. Using lumped-element resonators with inductor wires oriented relative to an in-plane magnetic field, we directly probe the superfluid stiffness via the kinetic inductance. As the field increases, the resonance frequency exhibits a non-monotonic and strongly anisotropic evolution that cannot be explained by orbital pair breaking alone. We show that this behavior is consistent with the emergence of Bogoliubov Fermi surfaces, which selectively suppress the supercurrent response depending on the direction of the magnetic field. Microscopic calculations of the stiffness tensor capture the observed anisotropy driven by the interplay of Zeeman and orbital Fulde-Ferrell effects. Our results establish microwave stiffness measurements as a sensitive probe of anisotropic gapless superconductivity in hybrid systems.
\end{abstract}

\maketitle

Bogoliubov Fermi surfaces (BFS) were originally predicted in certain centrosymmetric superconductors with unconventional pairings \cite{Agterberg2017Mar}, inherently protected by inversion symmetry. However, Yuan and Fu proposed in 2018 that ``banana-shaped'' BFS could be induced in a proximitized two-dimensional electron gas (2DEG) with Rashba spin-orbit coupling (SOC) in the presence of an in-plane Zeeman field \cite{Yuan2018gapless}. These BFS correspond to zero-energy Bogoliubov quasiparticles, coherent superpositions of electrons and holes, residing on two isolated segments on opposite sides of the original Fermi surface. They constitute the signature of a gapless superconducting phase that is induced by an in-plane magnetic field and stabilized by the resilience of the parent superconductor. This concept gained significant experimental traction with the discovery of ``segmented Fermi surfaces'' induced by finite Cooper pair momentum in a bilayer of Bi$_2$Te$_3$ on NbSe$_2$ \cite{Zhu2021}. That milestone revitalized interest in Fulde-Ferrell (FF) physics \cite{Fulde_Ferrell_1964,Larkin1964,Fulde1965,Fulde1969}, demonstrating that BFS stability can be mediated by an in-plane magnetic field penetrating the interface between the parent superconductor and the 2D layer. This leads to an orbital Fulde-Ferrell (OFF) type of phase bearing similarities to those recently explored in layered transition metal dichalcogenides and Moiré systems \cite{Wan2023Jul, Xie2023Jul, Cho2025, Zhu2025Jul, Cao2026}. Alongside these discoveries, theoretical efforts have intensely focused on emergent phases driven by in-plane magnetic fields and finite momentum superconductivity \cite{YuanFu2021}, particularly as a consequence of the presence of BFS \cite{Papaj2021}.

\begin{figure*}[t!] 	
	\centering
	\includegraphics[width=1\textwidth]{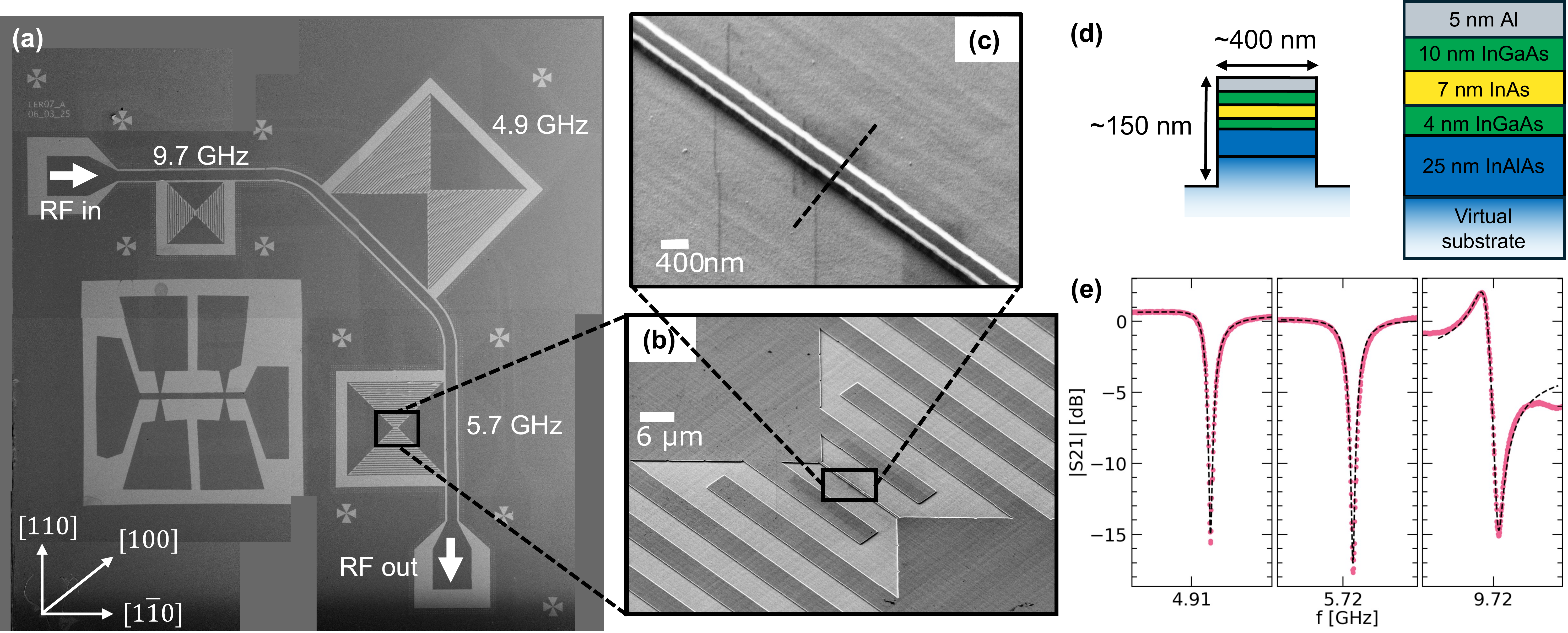} 
    \caption
	{\textbf{Sample overview:} \textbf{(a-c)} Three lumped-element resonators fabricated in Al/InAs with resonance frequencies at 4.9\ GHz, 5.7\ GHz, 9.7\ GHz are capacitively coupled to the same transmission line, used to measure the complex transmission coefficient $S_{21}$. The three devices are oriented along different crystal axes. A large Hall bar structure allows DC-access and on-chip heating. \textbf{(b)} Each resonator consists of an interdigitated capacitor shunted by a narrow wire acting as inductor. \textbf{(c)} The wires are typically 200-400\ nm wide and 20$\ \mu$m long. The resonance frequency is tuned with the number of fingers in the capacitor. The design allows a fine alignment with in-plane magnetic field and a simultaneous measurement of the dependence with the angle between the current density $\mathbf{J}$ and the field $\mathbf{B}$. \textbf{(d)} The devices are obtained by nano-patterning the hybrid Al/InAs heterostructure to form a 150\ nm mesa. The illustrative profile corresponds to the dashed line cut in \textbf{(c)}. \textbf{(e)} $|S_{21}|$ as a function of frequency showing the resonance curves for the three resonators. The frequency shift and the change in the quality factor allow monitoring the dependence of the kinetic inductance with magnetic field, which in turn provide access to the superfluid stiffness tensor.}
    \label{fig:design}
\end{figure*}

In this work, we present the first clear observations of a gapless phase characterized by the emergence of BFS in a hybrid superconductor/semiconductor platform. Our heterostructure consists of a 2DEG with strong SOC hosted by an InAs quantum well, proximitized by a thin epitaxial Al layer separated by a potential barrier \cite{Shabani2016,BaumgartnerPRL2021}. The interplay of SOC and Zeeman fields in proximitized materials has already proven to be a rich playground in the quest for topological superconductivity in both 1D wires \cite{Fu2008,Lutchyn2010,Oreg2010,Potter2012} and 2D planar junctions \cite{Pientka2017,Mayer2019}. More recently, these same ingredients have been shown to produce non-reciprocal effects in 2D Al/InAs heterostructures \cite{BaumgartnerNature2022}. As an alternative explanation, the superconducting diode effect and non-reciprocal superconductivity have been linked to orbital effects and finite momentum superconductivity \cite{YuanFu2022,Davydova2022,Davydova2024Nov}, including the case of planar Josephson junctions in Al/InAs \cite{Banerjee2023}. We demonstrate that the electrodynamics of these systems is most accurately described by a combination of both Zeeman and orbital mechanisms, opening the door to a much richer phase diagram.

Motivated by recent efforts to analyze the effective pairing in hybrid superconductor/semiconductor heterostructures via their electrodynamic response \cite{Phan2022BFS,Babkin2024BFS}, we utilize microwave lumped-element resonators to directly measure the superfluid stiffness as a thermodynamic observable. The emergence of Bogoliubov quasiparticles inherently impacts the superfluid stiffness and the dissipation, which can be precisely monitored in our resonant circuit. While previous work on distributed Al/InAs microwave resonators attributed anomalous frequency deviations to the emergence of BFS and anisotropic g-factor \cite{Phan2022BFS}, the effect of pinned vortices is hard to rule out \cite{Fuchs2022Nov}. Our lumped-element design allows us to probe the stiffness with high precision as a function of the in-plane magnetic field orientation. As the field increases, the response becomes strongly anisotropic and exhibits a non-monotonic dependence, a definitive signature of BFS selectively suppressing the supercurrent response. We support these observations with a complete microscopic theoretical analysis, calculating the stiffness tensor within a pure Zeeman/SOC scenario, a pure orbital Fulde-Ferrell scenario, and a combined model. We believe these combined mechanisms and experimental methods will be broadly applicable for interpreting recent and future experiments in layered superconducting devices \cite{Wan2023Jul,Xie2023Jul,Zhao2023Nov,Levichev2023Sep}.

We access the superconducting stiffness of the Al/InAs heterostructure via the kinetic inductance. As shown in Fig.~\ref{fig:design}(a-c) we fabricated three lumped-element resonators coupled capacitively to the same transmission line. We probe the transmission coefficient $S_{21}(f)$ of the line as a function of frequency, to find the corresponding resonances as in Fig.~\ref{fig:design}(e). Each resonator consists of an interdigitated capacitor shunted by a narrow wire acting as inductor. They are obtained by nano-patterning the heterostructure in Fig.~\ref{fig:design}(d), leaving a mesa of $\sim$150\ nm. The wires are typically 20\ $\mu$m long and 200-400\ nm wide. The resonance frequencies are changed by varying the length and number of fingers in the capacitor. The measured values at zero magnetic field are 4.9\ GHz, 5.7\ GHz and 9.7\ GHz for the resonators oriented at 45$^{\circ}$, 0$^{\circ}$ and 90$^{\circ}$ with respect to crystal [1$\bar{1}$0] direction, taken as the horizontal (see panel (a) of Fig.~\ref{fig:design}). The geometrical design allows a precise alignment with respect to an applied in-plane magnetic field. The three resonators yield simultaneous information regarding the relative orientation of the current density and the field. For every value of the in-plane field, we employ a robust and reproducible protocol, whose reliability we have verified through extensive consistency checks, such that the out-of-plane component of the magnetic field is $\lesssim 5-10 \ \mu$T  (see Supplementary Information). Even in the presence of a residual out-of-plane component, the narrowness of the wires prevents vortex nucleation inside the inductor. Additionally, artificial pinning sites were patterned in the ground plane nearby the measured structure to avoid dissipation due to vortex motion. 

The geometrical capacitance $C$ and inductance $L_g$ can be obtained by finite-element simulations and predict a resonance frequency $f_g=\left(2\pi \sqrt{L_gC}\right)^{-1}$ (see Supplementary Information). The measured frequency is lowered by the contribution of the kinetic inductance $L_k$, such that $f_0=\left(2\pi \sqrt{(L_g+L_k)C}\right)^{-1}$. The kinetic energy stored in the supercurrent of a wire oriented along a principal axis $i$, with length $\ell$ and cross-sectional area $S$, defines the kinetic inductance $L_{k,i} = \left(\hbar/2e\right)^2 \left(\ell/S\right) D_{ii}^{-1}$, where $D_{ii}$ is the diagonal component of the superfluid stiffness tensor $D_{ij}$ \cite{schmidt2013physics,tinkham2004introduction}. The macroscopic electrodynamic response of the system can be linked to its microscopic properties via the gauge-invariant phase gradient of the superconducting condensate. In the presence of a vector potential $\mathbf{A}$, the free energy density of the condensate expands to second order as
\begin{equation}
\mathcal{F} = \mathcal{F}_0 + \frac{1}{2} \sum_{i,j} D_{ij} K_i K_j,
\label{eq:free_energy}    
\end{equation}
where $\mathbf{K} = \nabla \phi - (2e/\hbar)\mathbf{A}$. The supercurrent density $\mathbf{J}$ is the thermodynamic conjugate to the vector potential, $J_i = -\partial \mathcal{F} / \partial A_i$. Assuming $\nabla \phi = 0$, this yields the London equation
\begin{equation}
J_i = -\left(\frac{2e}{\hbar}\right)^2 \sum_j D_{ij} A_j.
\label{eq:london}    
\end{equation}

In our experiment, we are sensitive to the component of the stiffness $D_{ii}$ related to the supercurrent density oscillating in the wire in response to the probing microwave field. Taking the $x$-direction as that oriented along the wire, we directly refer to $D\equiv D_{xx}$. It is physically synonymous with the superfluid density $n_{S}\equiv ({4m}/\hbar^2)D$. The stiffness is probed directly via the kinetic inductance $L_k$ of the lumped-element resonators. Thus, the shift in the microwave resonance frequency
\begin{equation}
\frac{\Delta f_0}{f_0} = \frac{\alpha}{2}\frac{\Delta D}{D}=\frac{\alpha}{2}\frac{\Delta n_{S}}{n_{S}},
\label{eq:relative_shift}
\end{equation}
serves as a direct, thermodynamic probe of the diagonal components of the stiffness tensor $D_{xx}(B)$. Here $\alpha=L_k/L$ is the kinetic inductance fraction. To calculate $D_{ij}$ microscopically, we exploit exact gauge equivalence. A uniform vector potential $\mathbf{A}$ is mathematically equivalent to a spatial phase twist in the order parameter, $\Delta(\mathbf{r}) = \Delta_0 e^{i 2\mathbf{q} \cdot \mathbf{r}}$, where 2$\mathbf{q}$ is the Cooper pair momentum \cite{Scalapino1993Apr}. Under this transformation, the stiffness tensor is determined entirely by the curvature of the free-energy with respect to this momentum
\begin{equation}
D_{ij}(B) = \frac{1}{4} \left. \frac{\partial^2 \mathcal{F}}{\partial q_i \partial q_j} \right\vert{}_{\mathbf{q} = \mathbf{q}_0},
\label{eq:stiffness}    
\end{equation}
evaluated at the minimum $\mathbf{q}_0$, which guarantees zero net ground-state current. 

\begin{figure*}[] 	
	\centering
	\includegraphics[width=1\textwidth]{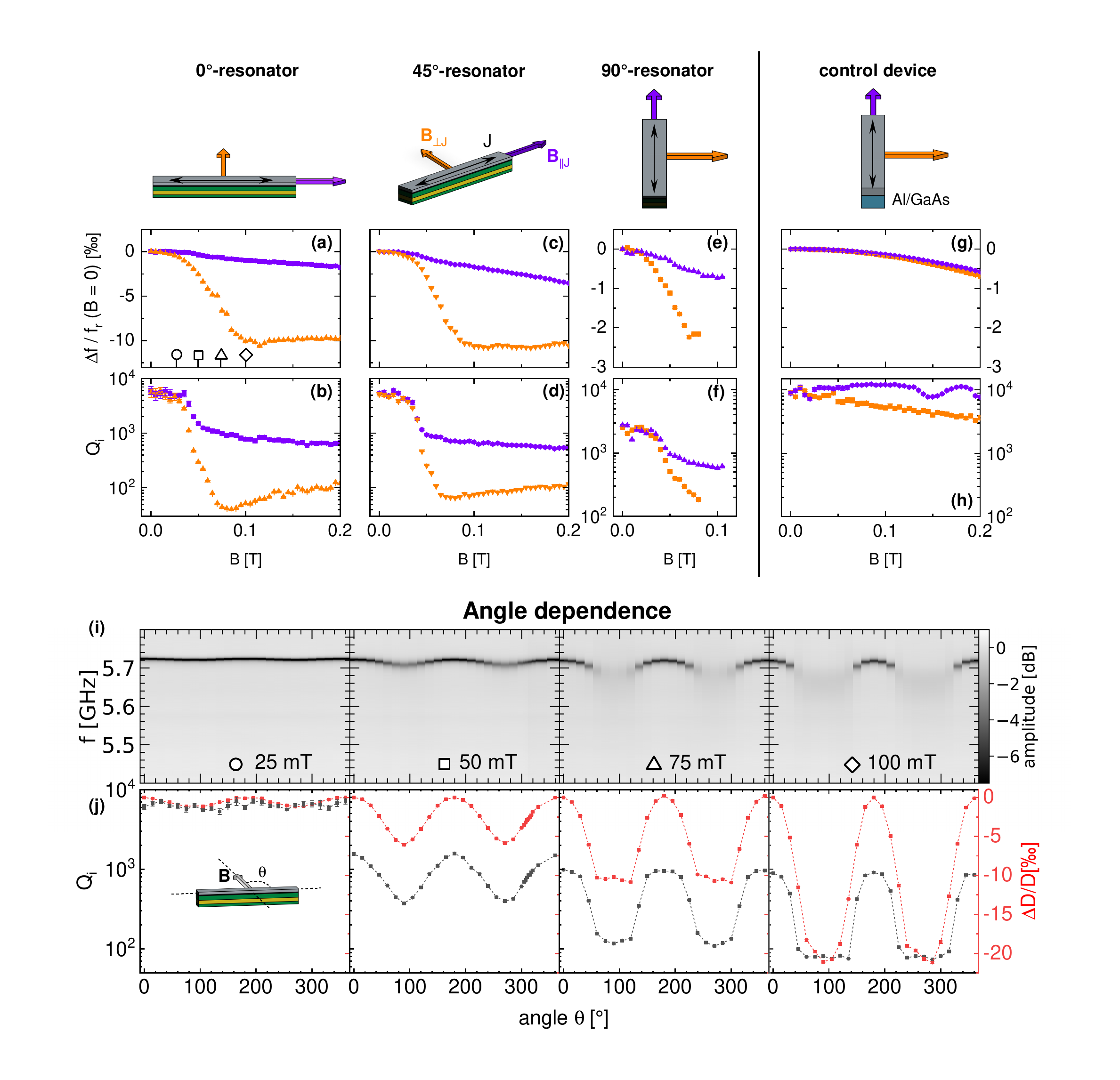} 
    \caption{\textbf{Response with in-plane field:} for all three devices oriented at \textbf{(a-b)} 0$^{\circ}$, \textbf{(c-d)} 45$^{\circ}$ and \textbf{(e-f)} 90$^{\circ}$, as described in Fig.~\ref{fig:design}. The relative frequency shift is shown in the top panels \textbf{(a,c,e)} and the internal quality factor in the bottom panels \textbf{(b,d,f)} as a function of in-plane magnetic field applied parallel (purple line) and perpendicular (orange line) to each resonator. The relative orientation between the inductor wire and the applied field is indicated with a little sketch for each device. Following the same color-code, in \textbf{(g-h)} the relative frequency shift and internal quality factor as a function of magnetic field for a witness resonator made out of Al on a GaAs substrate. \textbf{(i,j)} Angle dependence for the 0$^{\circ}$ resonator taken rotating the field from being parallel ($\theta=0^{\circ}$) at $|B|=$25, 50, 75 and 100\ mT, as indicated by the symbols. \textbf{(i)} Reconstructed $|S_{21}|$ map as a function of probing frequency and $\theta$. \textbf{(j)} Extracted quality factor (black symbols) and relative superfluid stiffness shift (red symbols) as a function of $\theta$.}
    \label{fig:ns_shfit}
\end{figure*}

To extract the field-dependent kinetic inductance, we fit the resonance curves (Fig.~\ref{fig:design}(e)) to obtain the  resonant frequency and the internal quality factor. Figure~\ref{fig:ns_shfit} shows both the relative frequency shift (a,c,e) and the quality factor change (b,d,f) as a function of applied in-plane field for all the three resonators described in Fig.~\ref{fig:design}. In the case of the 0$^{\circ}$ (a,b) and 45$^{\circ}$ resonators (c,d), we were able to track the signal up to larger fields as compared to the 90$^{\circ}$ resonator (e,f). To facilitate readability, three sketches indicate the relative orientation of the wires with respect to the applied field. The purple (orange) arrow corresponds to the field being parallel (perpendicular) in each case. In all three resonators, a clear change of behavior is observed in the quality factor at $|B|\sim$ 50\ mT, pointing to the opening of an additional dissipation channel. The impact on the resonance frequency, which encodes the effect on the superfluid stiffness, depends on the angle between the field ${\bf B}$ and the current ${\bf J }$ in the inductor. The effect is stronger when ${\bf B} \perp {\bf J}$ (orange curves). 

To disentangle BFS-related effects from trivial pair-breaking features we measured a control sample with exactly the same geometry. The microwave resonators were fabricated using 8\ nm-thick epitaxially grown Al on a GaAs substrate. We observed only a weak $B^2$ dependence in all cases, as expected from orbital depairing (see Supplementary Information). An example is presented in Figure~\ref{fig:ns_shfit}(g,h) following the same color code. Clearly for the Al/InAs sample, the strong anisotropic suppression observed in Fig.~\ref{fig:ns_shfit}(a-f) for in-plane field perpendicular to the current cannot be attributed to pair-breaking effects. As discussed below, such non-monotonic and anisotropic behavior is a fingerprint of the emergence of BFS.

Further signatures of the BFS are observed in the angular dependence of the resonances for different in-plane field strengths. The four panels in Fig.~\ref{fig:ns_shfit}(i) show reconstructed maps of $|S_{21}|$ as a function of probing frequency and the angle $\theta$ formed by the field and the 0$^{\circ}$ resonator (see the sketch in the figure). The panels correspond to the values of $|B|$ indicated with symbols in (a). The extracted relative superfluid stiffness change $\Delta D/D$ (red symbols) and internal quality factor (black symbols) as a function of $\theta$ are shown in the four panels of Fig.~\ref{fig:ns_shfit}(j). For 25\ mT, where BFS are still not formed, $\Delta D/D$ and $Q_i$ exhibit a weak angular dependence which can be attributed to the impact of spin-orbit in  pair breaking (see Supplementary Information). Since the area of the BFS gets larger for larger fields, the range of angles where the superfluid stiffness is affected becomes also larger from 50\ mT to 100\ mT. $\Delta D$ deviates from the sinusoidal shape above 50\ mT and the quality factor follows the collapse of $D$ due to the increase in dissipation close to 90$^{\circ}$. The maximum suppression of the superfluid stiffness occurs around $\theta=$90$^{\circ}$ and 270$^{\circ}$. This change with the angle due to the enlarging of the BFS is emphasized by the corresponding $S_\text{21}(f,\theta)$ map in Figure~\ref{fig:ns_shfit}(i).

\begin{figure*}[t!]
    \centering
    \includegraphics[width=2\columnwidth]{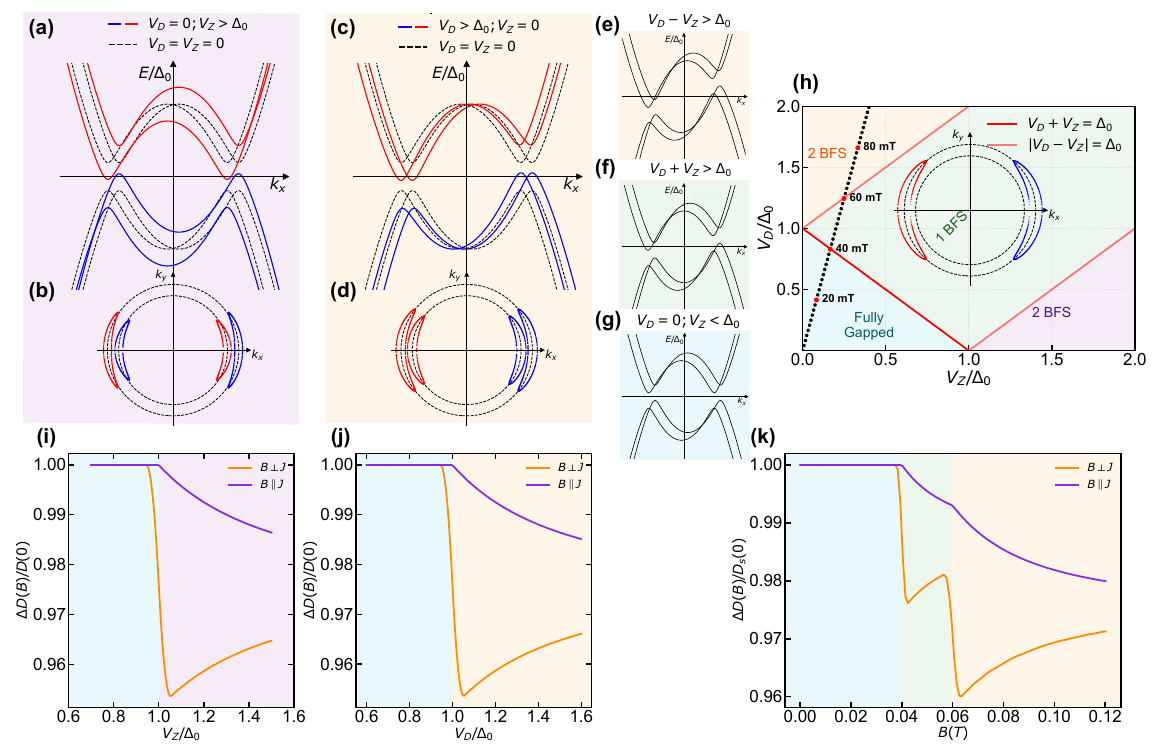}
    \caption{\textbf{Bogoliubov Fermi Surfaces:} \textbf{(a,c)} Energy bands for the BdG Hamiltonian obtained in a simplified model of the proximitized 2DEG with SOC. Dashed lines correspond to zero field. Continuous lines show the effect of \textbf{(a)} Zeeman energy $V_Z$ and \textbf{(c)} Doppler energy $V_D$ larger than the induced gap $\Delta_0$. In \textbf{(a)} electron-like (red) and hole-like (blue) bands shift with applied field (here in the $y$-direction) crossing the Fermi level, while in \textbf{(c)} they are tilted. \textbf{(b,d)} As a consequence ``banana-shaped'' BFS emerge in both cases along the $k$-direction perpendicular to the field. \textbf{(e-g)} A combination of both effects ($V_Z,V_D$) results in the successive transitions: \textbf{(g)} gapped phase $\rightarrow$ \textbf{(f)} 1-BFS $\rightarrow$ \textbf{(e)} 2-BFS as summarized in the phase diagram \textbf{(h)}. The 1-BFS pair shown inside the green region corresponds to the bands crossing the Fermi level in \textbf{(f)}. For the phase diagram, we use $V_Z=\mu_B g B/2 =$0.03$g B$ [meV/T] and $V_D=\eta_D e d v_F B \approx 10 \eta_D B$ [meV/T], with $v_F=$6.65 10$^{5}$\ m/s and $d=$15\ nm. The line is drawn for $g=$11.5, $\eta_D=$0.166 and $\Delta_0=$0.08\ meV. \textbf{(i-k)} Relative stiffness shift obtained from the curvature of the free energy along the $x$ direction calculated at zero temperature as a function of in-plane field applied parallel (purple) and perpendicular (orange) to $x$. \textbf{(i)} In the Zeeman Scenario the phase diagram is traversed horizontally by increasing $V_Z$ at $V_D=0$, while in \textbf{(j)} it is crossed vertically, i.e. increasing $V_D$ with $V_Z=0$. In \textbf{(k)} the plot is shown as a function of magnetic field following the path indicated with dots in \textbf{(h)}.}
    \label{fig:BFS}
\end{figure*}

In the scenario discussed by Yuan and Fu \cite{Yuan2018gapless}, the 2DEG is described using Rashba-split parabolic bands. In the Bogoliubov-de Gennes (BdG) picture, as a consequence of induced superconductivity, a gap $\Delta_0$ appears in the spectrum due to the coupling of electron and hole states at the Fermi level (dashed lines in Fig.~\ref{fig:BFS}(a,c)). In the presence of a magnetic field, the Zeeman energy shifts the bands up and down by an amount $\pm V_Z$, with $V_Z=g \mu_B B/2 $, proportional to the field strength $B$. Here, $g$ is the gyromagnetic factor of the 2DEG and $\mu_B$ is Bohr's magneton. When $V_Z>\Delta_0$, the gap closes as shown in (a) for the electron- (red) and hole-like (blue) bands. SOC prevents the gap from closing across the entire Fermi surface, and instead ``banana-shaped'' BFS form as shown in (b). They form in the direction of $k$-space perpendicular to the applied magnetic field (in the figure $\mathbf{B}=B\hat{y}$). 

In the orbital FF scenario (c-d), the penetration of the field in the interface between the parent superconductor (the Al layer) and the 2DEG induces circulating currents (proportional to the vector potential) which can be associated with momenta $\hbar \mathbf{q}$ and $\hbar \mathbf{q}_D$, in both layers respectively. Within this model of the system as composed of two 2D layers separated by the effective distance $d$, the Doppler momentum in InAs is $\hbar \mathbf{q}_D = e d (\mathbf{\hat{z}} \times \mathbf{B})$, proportional to the field strength $B$, and perpendicular to it. In the Al layer, the effect of the finite momentum $\hbar \mathbf{q}$ can be taken into account in the pairing $\Delta(\mathbf{r}) = \Delta_0 e^{i 2\mathbf{q} \cdot \mathbf{r}}$. In the 2DEG, there are two contributions: first, the induced pairing carries the information of the finite momentum of the parent superconductor. Second, the Doppler momentum contributes in the dispersion relation with an energy term $V_D=\hbar \mathbf{v}_F \cdot \mathbf{q}_D$. The effect of this term is to tilt the bands. When $V_D>\Delta_0$, the bands cross the Fermi level and BFS form also perpendicular to the applied field. In the real device, the supercurrents that develop to screen the magnetic field penetrating the heterostructure are reduced by a factor $\eta_D$, such that $V_D=\eta_D e v_F d B$. This factor is adjusted to fit the critical fields in the experiment as discussed below.

The scenarios presented in (a-b) and (c-d) correspond to only two possible paths in a rich phase-diagram (see (h)). An experiment most likely involves an interplay of both mechanisms: Zeeman and orbital FF. To illustrate this in (g) we show the bands in the gapped phase, where Zeeman energy is not enough to close the gap. The tilt induced by adding the Doppler contribution (f) produces first a single BFS pair (see inset in (h)) and then two (e).

The gapless phases will be prone to an anisotropic electrodynamic response. The stiffness tensor can be computed from Eq.~\eqref{eq:stiffness}, where the total free-energy must include the contribution of the Al and the 2DEG layers. At zero temperature, we need to calculate the ground-state energy as a function of $\mathbf{q}=q \hat{x}$. In the case of the Al layer, neglecting the effect of Zeeman energy due to the small g-factor yields $E_{\text{GS}}^{\text{Al}}(q)=E_{\text{GS}}^{\text{Al}}(0)+2 D^{\text{Al}}q^2$. For the 2DEG, in the Zeeman scenario (Fig.~\ref{fig:BFS}(a,b)) the calculation for $V_Z<\Delta_0$ yields a quadratic dependence as well $E_{\text{GS}}^{\text{2D}}({q})=E_{\text{GS}}^{\text{2D}}(0)+2D^{\text{2D}}q^2$. Within the gapped phase, the minimum of the total free energy is at $q=0$. Once $V_Z>\Delta_0$, the contribution of the bands associated to the BFS has to be subtructed from calculation of the ground-state energy of the 2DEG. It can be shown that this leads to a sudden change of curvature at $q=0$ from plus to minus $D^{\text{2D}}$, converting the parabolic dependence ($V_Z<\Delta_0$) into an asymmetric double-well (see Supplementary Information) \footnote{This asymmetric $q$-dependence is linked to non-reciprocal effects.}. Since $D^{\text{Al}}\gg D^{\text{2D}}$, the minimum is still very close, but not equal to zero. Finite momentum in the Al is necessary to stabilize the gapless phase in the 2DEG. The relative shift of the curvature of $E_{\text{GS}}^{\text{2D}}({q})$
calculated with the field parallel (purple, $\mathbf{B}=B\hat{x}$) and perpendicular (orange, $\mathbf{B}=B\hat{y}$) to the $D_{xx}^{\text{2D}}$ component of the tensor (i.e., parallel and perpendicular to the current density) are shown in Fig.~\ref{fig:BFS}(i). Evaluated very close to the origin, $D$ abruptly drops (due to the sign change of the 2DEG contribution) and as the field increases further, the free-energy becomes shallower. The hook-like feature is a consequence of this sudden drop. The effect of temperature is to wash out the abrupt transition (see Supplementary Information).

\begin{figure}[] 	
	\centering
	\includegraphics[width=0.5\textwidth, trim =.9cm 0cm 0cm 0cm, clip]{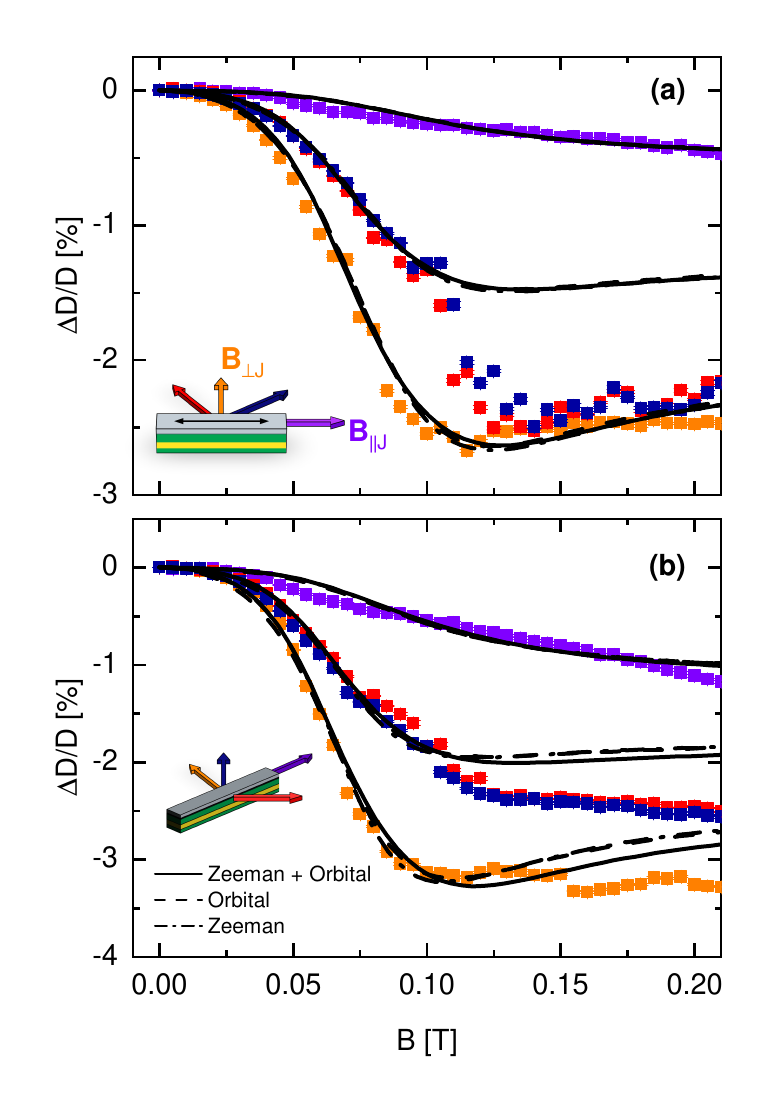} 
    \caption{\textbf{Comparison with theory:} for the resonators oriented at \textbf{(a)} 0$^{\circ}$ and \textbf{(b)} 45$^{\circ}$, as described in Fig.~\ref{fig:design}. Relative stiffness shift as a function of in-plane magnetic field applied parallel (purple), perpendicular (orange) and forming 45$^{\circ}$ (blue) and 135$^{\circ}$ (red) to each resonator. The relative orientation between the inductor wire and the applied field is indicated with a little sketch for each device. The experimental results are shown with dots while the lines correspond to the theory calculations: continuous line is the combined Zeeman/Orbital scenario, dashed line corresponds to Orbital-alone scenario and dot-dashed to Zeeman-alone scenario (see Fig.~\ref{fig:BFS} and discussion on parameters in text).}
    \label{fig:fit}
\end{figure}

In the orbital FF scenario (Fig.~\ref{fig:BFS}(c,d)), the ground-state energy contribution of the 2DEG for $V_D<\Delta_0$ is a horizontally shifted parabola $E_{\text{GS}}^{\text{2D}}(\mathbf{q})=E_{\text{GS}}^{\text{2D}}(0)+2D^{\text{2D}} |\mathbf{q}+\mathbf{q}_D|^2$. We see here that $E_{\text{GS}}^{\text{tot}}(\mathbf{q})\sim E_{\text{GS}}^{\text{tot}}(0)+ 2D^{\text{2D}}|\mathbf{q}+\mathbf{q}_D|^2 + 2D^{\text{Al}}|\mathbf{q}|^2$, will be minimal for $\mathbf{q}_0=-\mathbf{q}_D \;{D^{\text{2D}}/(D^{\text{2D}}+D^{\text{Al}}} ) $, which is $|\mathbf{q}_0|\ll |\mathbf{q}_D|$. This is reasonable because in order to have a zero ground-state net current, the fewer carriers in the 2DEG have to move faster than those in the Al. As soon as the BFS emerge, there is a negative contribution to the ground-state energy which results in the orange curve shown in Figure~\ref{fig:fit}(j). It is remarkable that although the mechanisms are very different, in both cases the response presents the hook-like feature and the gapless phase is stabilized by finite momentum in the Al. This is mostly due to the fact that the change of behavior is caused by the BFS which are very similar in both scenarios (b,d). In the most general case which corresponds to the path indicated with dots in Fig.~\ref{fig:BFS}(h), the stiffness shown in Fig.~\ref{fig:BFS}(k) presents two abrupt drops (orange curve) as a function of magnetic field accounting for the successive transitions across the gapped $\rightarrow$ 1-BFS $\rightarrow$ 2-BFS phases.  

Figure~\ref{fig:fit} presents the fits of the experimental stiffness data using realistic parameters for the Al/InAs heterostructure. To model the system accurately, we extract the material parameters from literature based on independent cyclotron resonance measurements on similar heterostructures \cite{BaumgartnerPRL2021,BaumgartnerNature2022,Chauhan2022}. We use an effective mass for the InAs 2DEG of $m=$0.0403\ $m_e$ and a carrier density of $n_{\text{2D}}=$8.5 $\times $10$^{11}\ \text{cm}^{-2}$, yielding a chemical potential $\mu=$50.6\ meV and a Fermi velocity $v_F=$6.65 $\times$ 10$^5$\ m/s. The spin-orbit coupling is set to $\lambda_{\text{SOC}}=$15\ $\text{meV nm}$ and $\Delta_0=$0.08$\ \text{meV}$, taken from a separate experiment on this material platform. Because the chemical potential is more than 600 times larger than $\Delta_0$, the physical bands are significantly steeper than those depicted in the illustrative presentation of Fig.~\ref{fig:BFS}, resulting in BFS that are much narrower features in $k$-space.

With these parameters the Doppler energy results $V_D(B)=\eta_D e v_F d B \approx 10\ \eta_D B$\ [meV/T] using $d=$15\ nm as distance between the layers. $\eta_D=1$ would correspond strictly to the ballistic limit. The Zeeman energy is $V_Z(B)= g \mu_B B/2 \approx 0.03 g B$ [meV/T]. The critical field \footnote{Here critical field applies to the magnetic field at which the minimum of the free energy changes.} $B_c$ such that $V_Z(B_c)=\Delta_0$ or $V_D(B_c)=\Delta_0$ (see Fig.~\ref{fig:BFS}(i,j)), determines the transition to the gapless phase within each scenario. If both mechanisms are considered together (see Fig.~\ref{fig:BFS}(k)), the equations $V_D(B_{c1})+V_Z(B_{c1})=\Delta_0$ and $V_D(B_{c2})-V_Z(B_{c2})=\Delta_0$ have to be satisfied for the critical fields $B_{c1},B_{c2}$. Such crossings are indicated in the phase-diagram of Fig.~\ref{fig:BFS}(h) for $\eta_D=$0.166 and $g=$11.5 and correspond to $B_{c1},B_{c2}=$0.04, 0.06\ T. In practice, to fit the data we proceed as follows: for the Zeeman and orbital scenarios we generate a large set of curves $D(B)$ vs $B$, like Fig.~\ref{fig:BFS}(i,j), for different critical fields $B_{c}\in [0.04-0.15]\ $T and temperatures $T_{\text{eff}}\in[100-500]\ $mK. Temperature here is a parameter that controls the broadening of the transition due to thermal and microwave-induced quasiparticles (see Supplementary Information). In the combined scenario we generate curves like Fig.~\ref{fig:BFS}(k) and sweep over $B_{c1}$, $B_{c2}$ and $T_{\text{eff}}$. We scale every curve to minimize the error with the experimental data which provides the relative weight of the stiffness of the Al and the 2DEG: $\Delta D(B)/D(0)=\Delta D^{\text{2D}}(B)/(D^{\text{2D}}+D^{\text{Al}})(0)$. Our best fit within each scenario is provided by the set of parameters that globally minimize the error.  

In Fig.~\ref{fig:fit} we compare three theoretical scenarios to the data in Fig.~\ref{fig:ns_shfit}. The results for the relative stiffness shift $\Delta D(B)/D(0)$ are shown for the resonators oriented at \textbf{(a)} 0$^{\circ}$ and \textbf{(b)} 45$^{\circ}$, as described in Fig.~\ref{fig:design}. For simplicity, in the following discussion we refer to the resonators with the labels a and b. We consider the cases where the in-plane magnetic field is applied parallel (purple), perpendicular (orange) and forming 45$^{\circ}$ (blue) and 135$^{\circ}$ (red) to each resonator (see the little sketches). Across all models, the analytic behavior perfectly captures the anisotropic dependence due to the emergence of BFS. When the field is applied parallel to the current density (purple curves), the BFS minimally affect the stiffness. Conversely, when the field is applied perpendicular to the current (orange curves), the emergence of BFS affects the current flow (see Fig.~\ref{fig:BFS}(b,d)), driving an abrupt suppression of the stiffness characterized by the distinct, non-monotonic hook-like dependence. In the intermediate 45$^{\circ}$ (blue) and 135$^{\circ}$ (red) cases, the response along the wire's direction corresponds approximately to the average of the main tensor components ($\tiltedparallel$ and $\perp$ to the field). For purely Zeeman-driven BFS, we find that an isotropic $g$-factor of $g=$35 (=39.5) for resonator a (b) yields a critical field of $B_c=$0.079$\ \text{T}$ (=0.07~T) and provides a good qualitative fit to the data (dot-dashed lines) for an effective temperature of $T_{\text{eff}}=$200$\ \text{mK}$, such that $k_B T_{\text{eff}}/\Delta_0=$0.215. For the orbital scenario alone (dashed lines) the best fit corresponds to $\eta_D=$0.101 (=0.114) for resonator a (b) with the same critical fields and effective temperatures. The combination of both mechanisms, which represents the most complete physical picture, yields also a similar fit (continuous line) with $B_{c1}=$0.068\ T (=0.063\ T) and $B_{c2}=$0.093\ T (=0.086\ T) for resonators a and b, respectively. With these values, $g$=5.47 (=5.88) and $\eta_D=$0.102 (=0.110) for an effective temperature $T_{\text{eff}}=$176$\ \text{mK}$. 

The analytical model successfully captures the distinct functional form of the anisotropic response. Within this simple model, both resonators a and b, yield similar values. Beyond this clean-limit theoretical framework, the experimental data for resonator a (b) reveals an anisotropy magnitude roughly a factor of four (two) larger than predicted for a pristine system. Rather than a simple discrepancy, this enhancement highlights the critical role of disorder in gapless superconducting phases. As recently established in Ref.~\cite{Babkin2024BFS}, the scattering of zero-energy Bogoliubov quasiparticles against the local disorder landscape provides a strong mechanism for further suppressing the stiffness. Crucially, this scattering is highly directional: when the in-plane field is applied perpendicular to the wire, the orientation of the supercurrent flow intersects the emergent Bogoliubov Fermi surfaces, rendering the macroscopic response exceptionally sensitive to momentum relaxation. To quantitatively account for this anisotropic scattering, we parameterize the disorder-induced momentum relaxation through a phenomenological anisotropy factor $\gamma$, defined by $D(B_{\perp}) = \gamma D(B_{\parallel})$ (see Supplementary Information). By incorporating this disorder-driven enhancement, the theoretical framework achieves excellent quantitative agreement with the experimental data, underscoring the interplay between emergent gapless states and mesoscopic disorder.

In summary, we have presented the first direct observation of an emergent gapless superconducting phase characterized by Bogoliubov Fermi surfaces in a proximitized hybrid 2DEG platform. By utilizing lumped-element microwave resonators, we intrinsically bypassed vortex-driven dissipation with the narrow inductor wires, allowing us to directly probe the highly anisotropic suppression of the superfluid stiffness under an in-plane magnetic field. The distinct, non-monotonic hook-like evolution of the kinetic inductance cannot be captured by standard pair-breaking mechanisms, but serves as a definitive macroscopic fingerprint of zero-energy Bogoliubov quasiparticles.

Through a rigorous microscopic analysis, we demonstrated that pure Zeeman spin-splitting, pure orbital Fulde-Ferrell and a combined theoretical framework qualitatively capture the non-monotonic anisotropic field-dependent evolution. Crucially, the stabilization of the gapless phase requires the presence of finite-momentum pairing. Furthermore, the striking enhancement of the measured anisotropy highlights the fundamental role that mesoscopic disorder plays in these gapless phases, providing highly directional momentum relaxation when the emergent Fermi surfaces align with the supercurrent \cite{RamshawKivelson2025}.

The experimental stabilization and direct measurement of this gapless phase carry profound implications for the understanding of topological quantum matter \cite{Flensberg2021, Yazdani2023hunting,DasSarma2023}. Most immediately, demonstrating the strong presence of orbital effects in Al/InAs heterostructures introduces critical new considerations into the quest for Majorana fermions in nanowires tailored from these platforms \cite{Aghaee2025Nature,Legg2026}. Beyond proximitized semiconductors, the interplay of finite-momentum pairing, spin-orbit coupling, and emergent Bogoliubov Fermi surfaces is increasingly recognized as a unifying thread across a broad frontier of modern condensed matter physics. Our results establish microwave stiffness as a powerful thermodynamic probe capable of uncovering similar unconventional and topological phases in strongly correlated layered materials \cite{Jin2025}, including transition metal dichalcogenides and twisted bilayer graphene \cite{Wan2023Jul,Xie2023Jul,Lin2022Moire,Zhu2025Jul}. Furthermore, understanding the precise dynamics of anisotropic momentum relaxation within these gapless states provides crucial microscopic insight into nonreciprocal transport phenomena, such as the superconducting diode effect \cite{Nadeem2023SDE,Davydova2024Nov,ShafferLevchenko2025}. As the search for topological superconductivity expands into novel platforms like altermagnets \cite{Ouassou2023Altermagnets}, where spin-split bands drive similar physics without macroscopic magnetization, such non-invasive electrodynamic techniques will be essential for mapping the complex phase diagrams of next-generation quantum materials.

\section*{acknowledgments}
The work in Regensburg was funded by the Deutsche Forschungsgemeinschaft (DFG, German Research Foundation) within Project-ID 314695032-SFB 1277-C03. N.P. and C.S. acknowledge funding by EU’s HORIZON-RIA Programme under Grant No. 101135240 (JOGATE). L.A. and L. T. are supported by CONICET, Argentina. L.T. acknowledges the Georg Forster Fellowship from the Humboldt Foundation. C.S acknowledges the DFG as part of the German Excellence Strategy – EXC3112/1 – 533767171 (Center for Chiral Electronics) and INST-89/561-1 FUGG. D.B. and C.S. acknowledge the Munich Quantum Valley program, which is supported by the Bavarian state government with funds from the Hightech Agenda Bavaria. 

\clearpage
\onecolumngrid 

\begin{center}
  \textbf{\large Supplementary Information: Emergence of Bogoliubov Fermi Surfaces in hybrid Al/InAs heterostructures}\\[.2cm]
\end{center}

\setcounter{equation}{0}
\setcounter{figure}{0}
\setcounter{table}{0}
\setcounter{page}{1}
\setcounter{section}{0}

\renewcommand{\theequation}{S\arabic{equation}}
\renewcommand{\thefigure}{S\arabic{figure}}
\renewcommand{\thetable}{S\arabic{table}}
\renewcommand{\thesection}{S\Roman{section}} 

In Sec.~\ref{ap:material} we describe the Al/InAs heterostructures used in the experiment. In Sec.~\ref{ap:device_fab} we present the details of the fabrication of the devices. In Sec.~\ref{ap:compensation} we discuss the method that we employ to compensate precisely any remaining out-of-plane component of the magnetic field. In Sec.~\ref{ap:Al_GaAs} we present the reference data corresponding to the same geometry of devices fabricated on Al on top of a GaAs substrate (without 2DEG). In Sec.~\ref{ap:crystal} we present a comparative figure to discuss the role of the crystal orientation. In Sec.~\ref{ap:temp_dep} we present additional data on the temperature dependence measured in different conditions for the Al/InAs devices of the main text. In Sec.~\ref{ap:model} we present the theoretical models that we use to calculate the response in different scenarios. In Sec.~\ref{ap:fitting} we discuss the fit of the experimental data.

\section{Al/InAs heterostructure}
\label{ap:material}
\begin{figure}[h!]
    \centering
\includegraphics[width=0.2\linewidth]{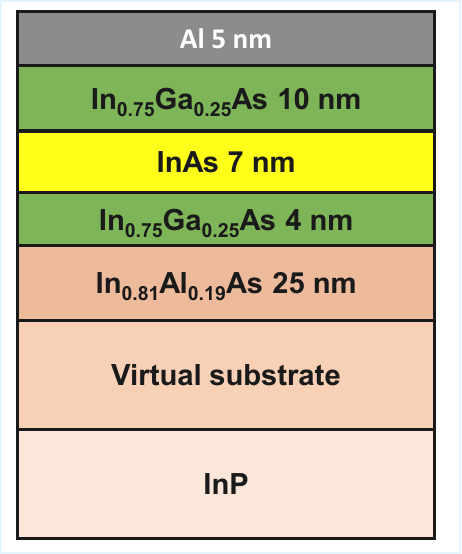}
    \caption{Detail of the Al/InAs Heterostructure \cite{Zhang2023_high_mob_QW}.}
    \label{fig:hetero}
\end{figure}

The material used for the experiment is a hybrid superconductor/semiconductor heterostructure based on an InAs quantum well \cite{Shabani2016}. The structure (see Fig.~\ref{fig:hetero}) is grown by Molecular Beam Epitaxy (MBE) resulting in a clean interface between the aluminum top layer and the underlying InAs 2DEG. The details of the growth can be found in Ref.~\cite{Zhang2023_high_mob_QW}. As a consequence of the clean interface the aluminum layer proximitizes the 2DEG resulting in induced superconductivity \cite{Shabani2016,Kjaergaard2016}. Additionally, the 2DEG incorporates a structural and bulk inversion asymmetry leading to Rashba and Dresselhaus SOC \cite{BaumgartnerNature2022}. Recent experiments found a ratio between Rashba and Dresselhaus SOC of around $10:1$ meaning that the influence of the crystal on the SOC should be rather small \cite{BaumgartnerNature2022}. The SOC impacts the spin texture of the bands, which is particularly important for the induced superconductivity in the presence of magnetic field. We designed an experiment for accessing information about the induced superconductivity in the 2DEG based on the kinetic inductance as a measurement observable.

To model the system accurately, we extract the fixed parameters from literature and independent cyclotron resonance measurements on similar heterostructures \cite{BaumgartnerPRL2021,BaumgartnerNature2022,Chauhan2022}. We use an effective mass for the InAs 2DEG of $m=$0.0403\ $m_e$ and a carrier density of $n_{\text{2D}}=$8.5 $\times $10$^{11}\ \text{cm}^{-2}$ \cite{Chauhan2022}, yielding a chemical potential $\mu=$50.6\ meV and a Fermi velocity $v_F=$6.65 $\times$ 10$^5$\ m/s. The spin-orbit coupling is set to $\lambda_{\text{SOC}}=$15\ $\text{meV nm}$ \cite{BaumgartnerPRL2021,BaumgartnerNature2022} and the induced gap $\Delta_0=$0.08$\ \text{meV}$, taken from a separate experiment on this material platform. We take $d=$15\ nm for the effective distance between the Al layer and the 2DEG.

\section{Device design and fabrication}
\label{ap:device_fab}
\subsection*{Design}
\begin{figure}[h!] 	
	\centering
	\includegraphics[width=0.6\textwidth]{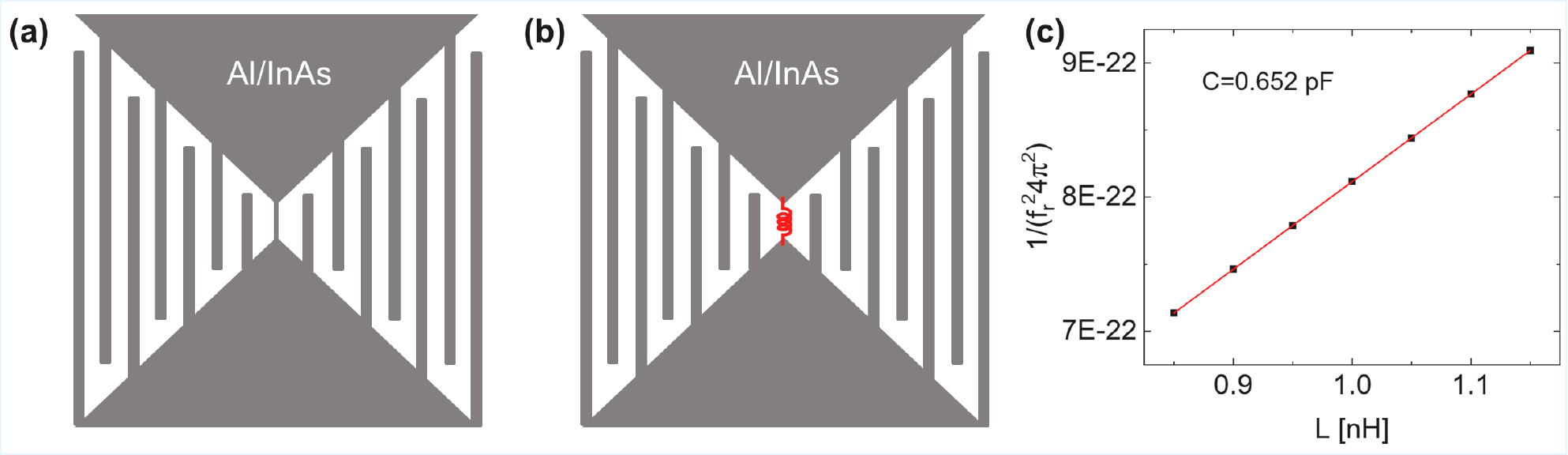} 
    \caption{(a) Lumped element resonator design. (b) Simulated geometry: The interdigitated capacitor is simulated by using a lossless metal layer with a sheet kinetic inductance of $10 \frac{\text{pH}}{\square}$. A substrate with a dielectric constant of $\varepsilon=12.4$ corresponding to $InP$ is used. The wire is replaced by an ideal inductor connecting both capacitor's electrodes. The inductance is varied during the simulations and the corresponding resonance frequency $f_r=\frac{1}{2\pi}\frac{1}{\sqrt{LC}}$ is extracted from the simulations. (c) $(2\pi f_{r})^{-2}$ as a function of the inductance of the ideal component. The capacitance of the geometry can be extracted by performing a linear fit.}
\label{fig:finite_element_simulations}
\end{figure}

We fabricated lumped element microwave resonators where the frequency shifts produced by changes in the kinetic inductance can be precisely measured \cite{annunziata2010tunableinductors}. Each resonator consists of an interdigitated capacitor shunted by a wire acting as inductor (see Fig.~\ref{fig:finite_element_simulations}(a)). The advantage of this geometry is that the inductors can be precisely aligned with the in-plane magnetic field and the crystal axis of the heterostructure.

For the design, we start with an estimate of the kinetic inductance per sheet $L_{\text{kin},\square}$ of the heterostructure, which can be obtained \cite{annunziata2010tunableinductors} from its sheet resistance $R_\square$:
\begin{align} \label{eq:kinetic_inductance}
    L_{\text{kin},\square}=\frac{\hbar R_\square}{1.76\pi k_\text{B}T_c}\approx 10 \text{ pH},
\end{align}
where $T_\text{c}=1.86$\ K is the critical temperature and $R_\square=13.54\Omega$ is the sheet resistance just before the transition, both measured using a Hall-bar fabricated in the same wafer. We then estimate the contribution of an inductor of length $L$ and width $W$ to the kinetic inductance of the lumped element resonators: $L_{\text{kin,induct}}=L_{\text{kin},\square}(L/W)$. We fix the length to 20 $\mu$m, then target a width of about 200\ nm. With these values the inductance is typically around 1\ nH. Making use of finite element analysis with \textit{Sonnet}, we obtain the the number of fingers of the interdigitated capacitor in order to reach resonance frequencies in the range of $4-8$ GHz. Since the resonances should be well separated in frequency, a different number of fingers was chosen for the three devices (see parameters summary in Table \ref{tab:resonators_params}). We perform the simulations using a metal layer with a sheet kinetic inductance of $L_{\text{kin},\square}\approx 10 \text{ pH}$.

The capacitance is extracted by replacing the thin wire in the \textit{Sonnet} simulation by an ideal inductor, sweeping its inductance in a realistic range and extracting the resulting resonance frequency (see Fig.~\ref{fig:finite_element_simulations}(b)). The capacitance is obtained from a linear fit of $(2\pi f_r)^2$ vs $L$ (panel (c)). Also, the best matching resonance frequency to the experiment gives information about the total inductance given by the sum of the geometric plus kinetic part. To deduce the pure geometric contribution, the same geometry can be evaluated by using a lossless metal layer without kinetic inductance ($L_{\text{kin},\square}=0$). For this, a thin wire without kinetic inductance connects both capacitor electrodes. The simulation of this structure returns the resonance frequency given by the geometric inductance alone. To calculate $L_{\text{geo}}$ the previously determined capacity is used (see table \ref{tab:resonators_params} column $L_{\mathrm{geo}}^{\mathrm{sim}}$):
\begin{align}
    L_{\text{geo}}=\frac{1}{f_\text{r}^2 4\pi^2 C}
\end{align}

In addition the same simulation can be performed with the sheet kinetic inductance $L_{\text{kin},\square}=10\ $pH added for the capacitors. This gives information about the parasitic inductance across the fingers of both electrodes (see table \ref{tab:resonators_params} column $L_{\mathrm{geo+par}}^{\mathrm{sim}}$). 

By comparing the measured resonance frequencies of all three resonators with the simulated resonance frequencies with the wire being replace by an ideal inductor one can obtain the total inductance of the wire compatible with the experiment (see table \ref{tab:resonators_params} column $L_{\mathrm{kin}}^{\mathrm{sim}}$).

\begin{table}[h!]
    \centering
    \caption{Summary of the resonators parameters.}
    \label{tab:resonators_params}
    \begin{tabular}{lccccccccccc}
        \toprule
        Orientation &
        Fingers &
        $C$ (pF) &
        Length ($\mu$m) &
        Width (nm) &
        $N_{\square}$ &
        $L_{\mathrm{kin}}$ (nH) &
        $f_r$ (GHz) &
        $Q_i$ &
        $L_{\mathrm{geo}}^{\mathrm{sim}}$ (nH) &
        $L_{\mathrm{geo+par}}^{\mathrm{sim}}$ (nH) &
        $L_{\mathrm{kin}}^{\mathrm{sim}}$ (nH)
        \\
        \midrule
        $0^{\circ}$  & 42  & $C_0=$ 0.652 & 21 & 200 & 105 & 1.05 & 5.7 & 6240 & 0.157 & 0.254 & 1.190\\
        $45^{\circ}$ & 62  & $C_{45}=$ 1.239 & 21 & 400 & 52.5 & 0.525 & 4.9 & 4727 & 0.211 & 0.307 & 0.853\\
        $90^{\circ}$ & 26 & $C_{90}=$ 0.289 & 21 & 200 & 105 & 1.05 & 9.7 & 2329 & 0.108 & 0.201 & 0.941 \\
        \bottomrule
    \end{tabular}
\end{table}

For one of the resonators, the 90$^{\circ}$ one, whose frequency is 9.7 \ GHz, there was a confusion in the e-beam lithography file which produced less fingers in the capacitor.

\subsection*{Fabrication}
Using a PMMA mask, the sample design was defined by e-beam lithography. Afterwards two separated wet chemical etching procedures removed the aluminum top layer and the semiconductor heterostructure down to the buffer layer. For the aluminum etching step a commercially available phosphor acid based solution called etchant type D (optimized for III-V semiconductors) by the company \textit{Transene} was used. The semiconductor layers were etched by a solution including citric acid, phosphoric acid and hydrogen peroxide following the same recipe as described in the supplementary material of \cite{BaumgartnerNature2022}. The sample was glued and bonded with aluminum wires to a home-built PCB.

\section{Measurement routine and magnetic field compensation} 
\label{ap:compensation}
\begin{figure}[h!] 	
	\centering
	\includegraphics[width=0.5\textwidth]{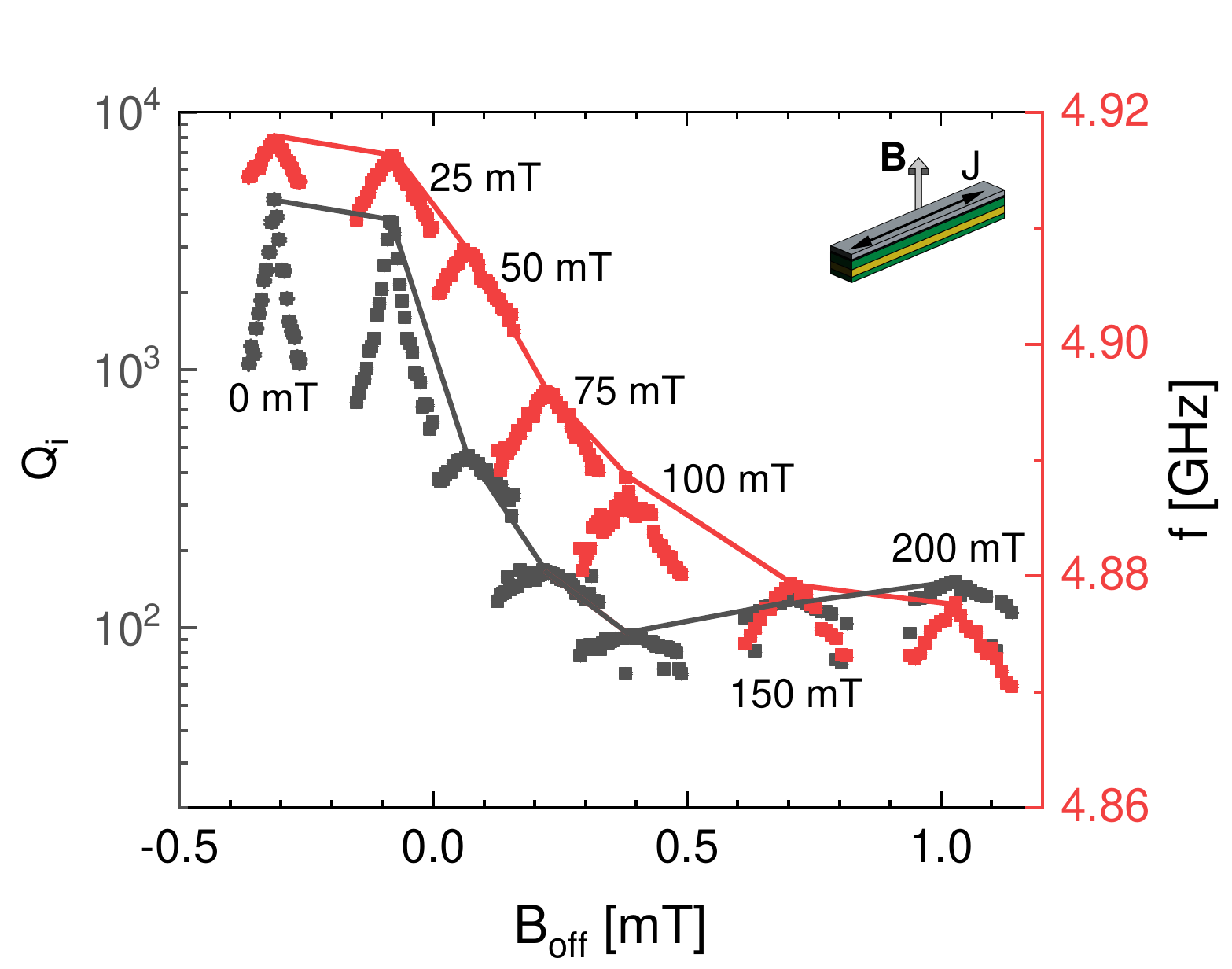}
        \caption{Frequency and quality factor as a function of $B_\perp$ for the 4.9\ GHz resonator, for selected in-plane field amplitudes applied at $45^{\circ}$ from the resonator corresponding to $\theta=90^{\circ}$ in the main text.
        The maxima of resonance frequency are connected by a guiding line to visualize the data used for calculating the blue curve plotted in Fig.~4(b) in the main text.} 
    \label{fig:compensation}
\end{figure}

The measurements were conducted in a dry dilution refrigerator including a 6-1-1 Tesla vector magnet. Detecting subtle changes in the superfluid density relies on a precise determination of the resonance frequency free from the influence of vortices induced by a finite out-of-plane field component. This component may originate from a misalignment of the sample with respect to the in-plane field. To make sure that the out-of-plane field component is as close to zero as possible, an out-of-plane field dependence of the resonance is measured for each in-plane field value. In addition, at every out-of-plane field the sample is heated above $T_\text{c}$ by applying a current $I>I_c$ to the Hall-bar structure depicted on the left lower corner of Fig.~1 in the main text. Using this procedure the sample can be heated up without a strong impact on the temperature of the mixing chamber and cold finger. After the heat pulse the sample cools down for about 45\ s and then $S_{\text{21}}(f)$ data for the corresponding out-of-plane plus in-plane field is measured (see Fig.~\ref{fig:compensation}). Therefore, each point is a field-cooling experiment, which should correspond to the thermodynamic equilibrium condition. The range for the sweeps of the out-of-plane field is selected by a $B_{\text{off}}(B_\parallel)$-function determined from a measurement with smaller precision and larger $B_\perp$-range. From this rough measurement the range needed to find the right $B_{\text{off}}$-value and time required per in-plane field is optimized.

From the $B_\perp$-dependence at each in-plane field the highest resonance frequency was extracted by fitting $S_{\text{21}}(f)$ using the circle fitting algorithm developed in Ref.~\cite{Probst2015S21}. These values are plotted as a function of the corresponding in-plane field in the main text (see e.g. Fig.~2). The compensation procedure of the out-of-plane field is illustrated in Fig.~\ref{fig:compensation} for the 4.9\ GHz resonator. The $B_\perp$-dependence is plotted for selected in-plane fields applied along the $\theta=90^{\circ}$ orientation. The extracted data correspond to the frequency shift used to calculate the blue curve in Fig.~4(b).

\section{Data for Al/GaAs reference device}
\label{ap:Al_GaAs}

In addition to excluding any vortex contribution in the inductance, we wanted to verify that the effect caused by the applied in-plane field was arising essentially from the proximitized 2DEG in the Al/InAs heterostructure. We fabricated devices with the same geometry using 10\ nm-thick Al epitaxially grown on top of an intrinsic GaAs substrate. We characterized these devices following the same procedure as for the Al/InAs samples.

\subsection*{B dependence}
\begin{figure}[h!] 	
    \centering
\includegraphics[width=\textwidth]{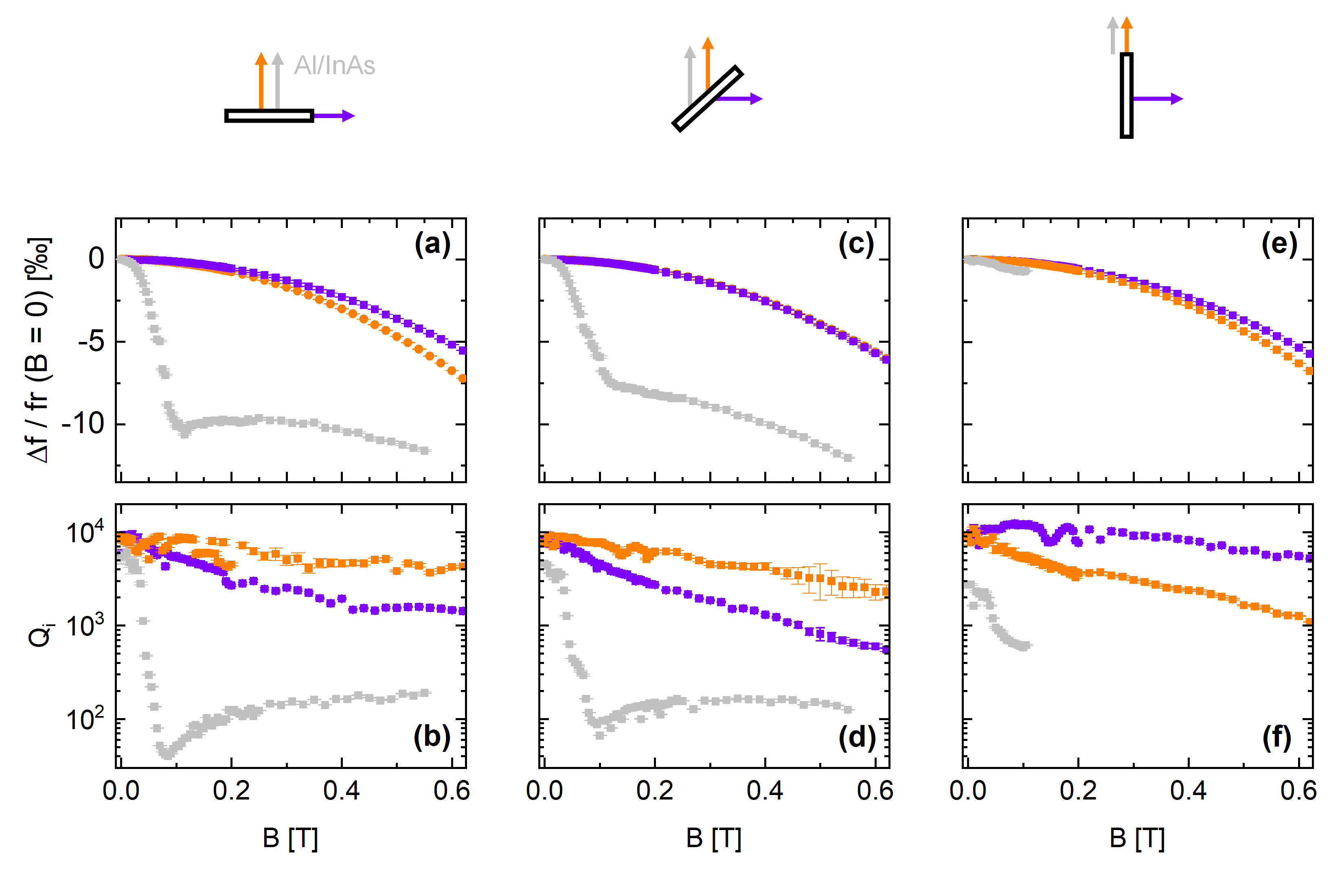} 
    \caption{\textbf{Magnetic field dependence of Al/GaAs resonators:}\textbf{(a,c,e)} Relative frequency shift and \textbf{(b,d,f)} internal quality factor measured for the three reference resonators. The orientations of the wires relative to the applied in-plane field are indicated in the sketches on top of each data set. For comparison, the grey line shows results for one in-plane field orientation of the Al/InAs device discussed in the main text.}
    \label{fig:AlGaAs_B}
\end{figure}

In Fig.~\ref{fig:AlGaAs_B} we show the relative frequency shift (a,c,e) and internal quality factor (b,d,f) as a function of applied in-plane field for the three resonators. In (a,b) we show the results for the $0^{\circ}$-resonator with the field applied parallel (purple line) and perpendicular (orange) to the inductor. For comparison, in gray we show in one of the cases, the result for the corresponding Al/InAs resonator (data of Figure~2). Panels (c,d) correspond to the 45$^{\circ}$ resonator and (e,f) to the 90$^{\circ}$ resonator, as indicated by the sketches. In every case for the Al/GaAs sample, we observe $\Delta f/f \propto -B^2$, as expected from orbital depairing \cite{tinkham2004introduction,annunziata2010tunableinductors,Splitthoff2022,muller2022magnetic}.

\subsection*{Angle dependence}
\begin{figure}[h!] 	
    \centering
    \includegraphics[width=0.9\textwidth]{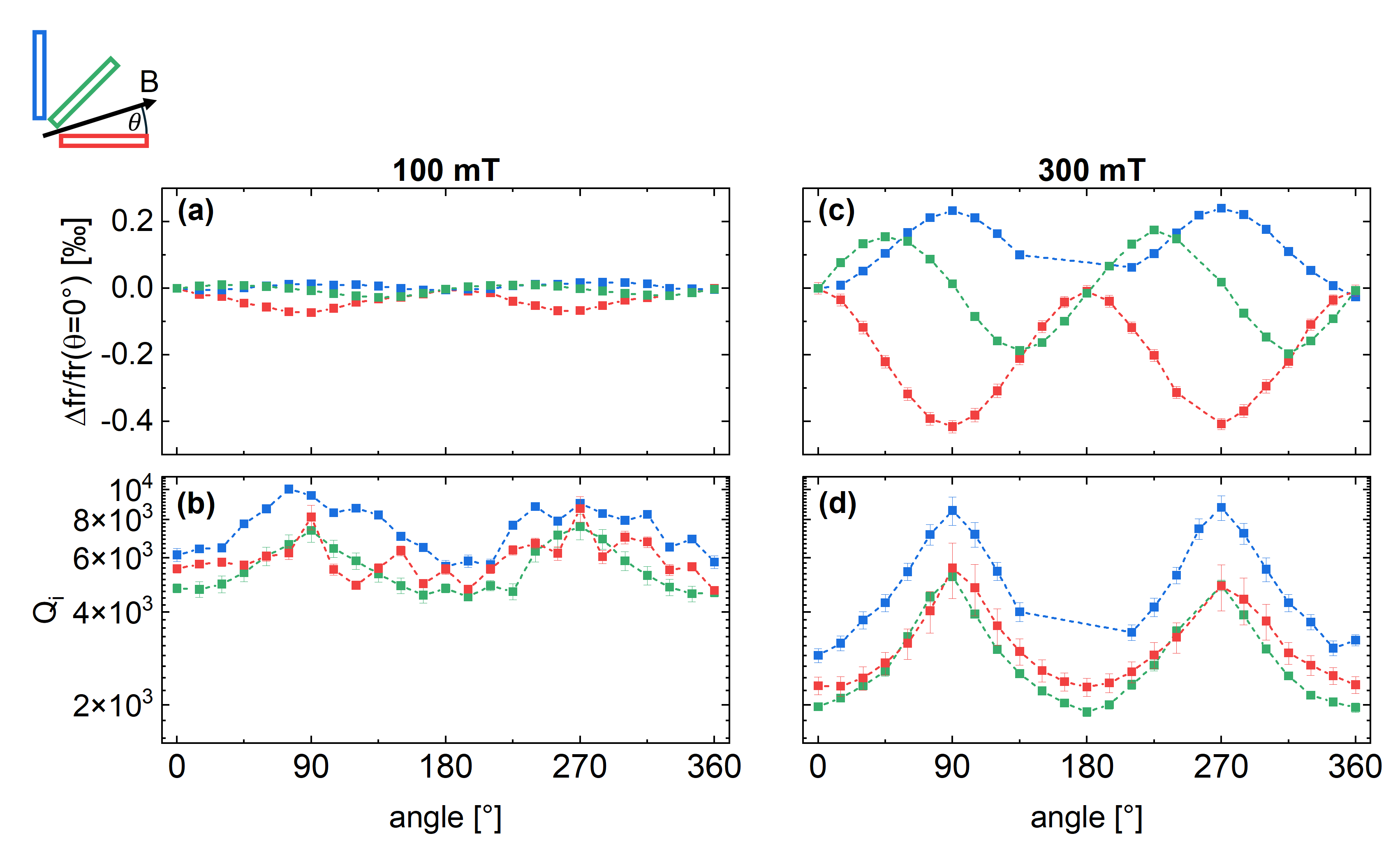} 
    \caption{\textbf{Angle dependence of Al/GaAs resonators:} \textbf{(a,c)} Angle dependence of the frequency shift and \textbf{(b,d)} the internal quality factor for the three different devices (red, green, blue) at in-plane fields of $100$\ mT and $300$\ mT.}
    \label{fig:angle_dependence_AlGaAs}
\end{figure}

In addition to the amplitude dependence for just two different in-plane field orientations, a full angle dependence was measured for $|B|=$100, 300\  mT. The results are shown in Fig.~\ref{fig:angle_dependence_AlGaAs}. The beauty of our design is that we can probe simultaneously three different relative orientations of wire and field (indicated with red, green and blue). The angle $\theta$ in the plane is measured with respect to the 0$^\circ$ resonator (in red). Fig.~\ref{fig:angle_dependence_AlGaAs}(a), shows a small angle dependence at 100\ mT, on the order of 0.1 \textperthousand, to be compared with the results presented in Fig.~2(i,j) in the main text. The strong angle dependence of the $Al/InAs$ resonators underlines the impact of the anisotropic closure of the gap caused by the presence of the emerging BFS.  

It is interesting that for the Al/GaAs sample, the angle dependence reveals a fixation of the small quality factor dependence to the crystal axis whereas the frequency shift is determined by the angle between field and current. The resonance frequency shift is stronger when the in-plane field is perpendicular to the current axis and weaker when it is parallel to the current. The quality factor on the other hand has clearly defined maxima at the field angles of $90^{\circ}$ and $270^{\circ}$ for all resonators. A possible explanation could be a competition between losses to the environment. Somehow, losses due to the coupling to two-level systems linked to the crystal orientation should be dominant in comparison to losses caused by depairing.

\section{Discussion on the role of Crystal Orientation}
\label{ap:crystal}

\begin{figure}[h!] 	
    \centering
\includegraphics[width=0.9\textwidth]{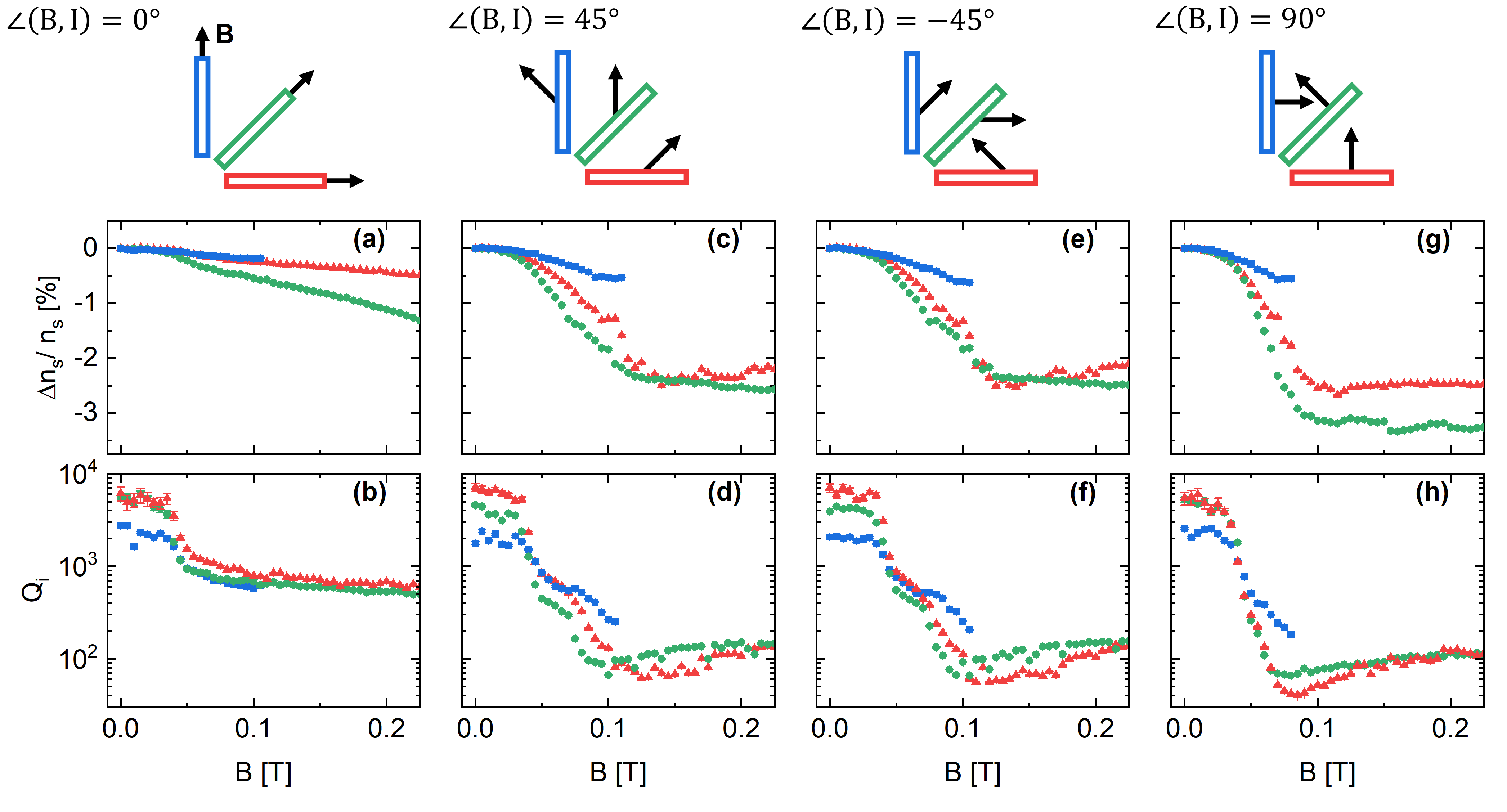} 
    \caption{\textbf{Comparison between different crystal orientations:} \textbf{(a,c,e,g)} Relative frequency shift and \textbf{(b,d,f,h)} internal quality factor as a function of in-plane field strength for the resonators fabricated along three different crystal orientations. Red corresponds to the $[1\overline{1}0]$-orientation, green to the $[100]$-orientation and blue to $[110]$-orientation. The sketch indicated the classification according to the angle between field and the current: \textbf{(a,b)}0$^\circ$,\textbf{(c,d)}45$^\circ$, \textbf{(e,f)}-45$^\circ$ and \textbf{(g,h)}90$^\circ$. Note that for the ac-current $135^{\circ}$ is the same as $-45^\circ$ for the $0^{\circ}$-resonator.}
    \label{fig:fr_current_to_field}
\end{figure}

As mentioned in the previous section, the design of the sample allows an immediate comparison between the frequency response of devices oriented along different crystal axis. For a better visualization of the effect of the crystal lattice on the in-plane field dependence the data in Fig.\ref{fig:fr_current_to_field} is sorted according to the angle between field and current.

Overall, there is a little effect of the crystal. The internal quality factor \textbf{(b,d,f,h)} shows very similar dependence independently of the crystal orientation (red, green, blue), dominated mainly by the anisotropy of the BFS. For the relative frequency shift, translated into relative superfluid density shift \textbf{(a,c,e,g)} there are small deviations. In all cases, the $[100]$-orientation (green) deviates more than the $[1\overline{1}0]$-orientation (red) and this, in turn, deviates more than the $[110]$-orientation (blue).    

\section{Additional data for Al/InAs}
\label{ap:temp_dep}
\subsection*{Temperature dependence}
\begin{figure}[h!] 	
    \centering
    \includegraphics[width=\textwidth]{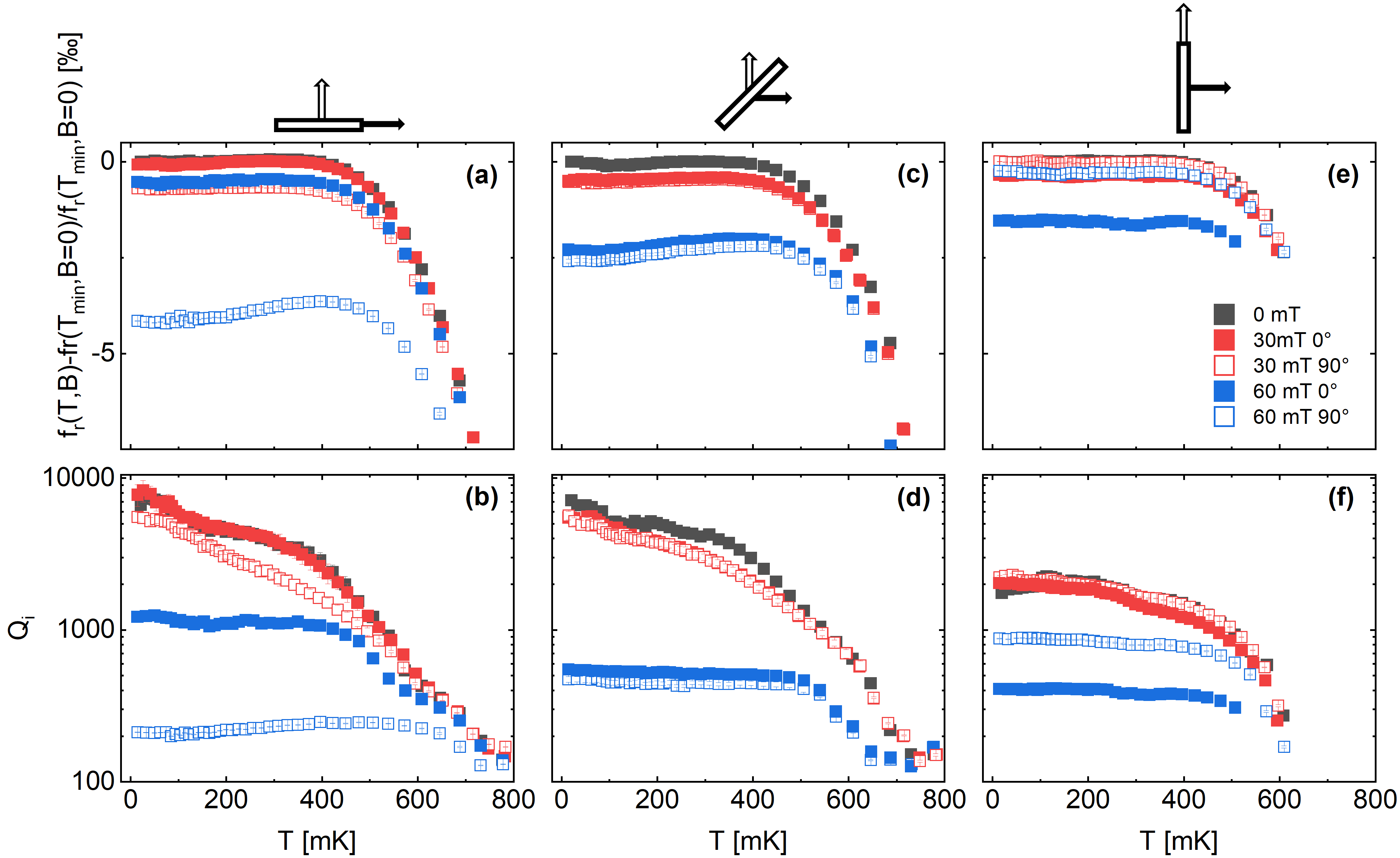} 
    \caption{\textbf{Temperature dependence:} \textbf{(a,c,e)} Relative frequency shift and \textbf{(b,d,f)} quality factor as a function of temperature for $B=$0 (black), 30\ mT (red) and 60\ mT (blue) applied parallel (full symbols) and perpendicular (empty symbols) to the $[1\overline{1}0]$-orientation. The sketches show the orientation of the resonators' wires relative to the applied in-plane field: \textbf{(a,b)} correspond to the 0$^{\circ}$-resonator, \textbf{(c,d)} to the 45$^{\circ}$-resonator and \textbf{(e,f)} to the 90$^{\circ}$-resonator. }
    \label{fig:Tdep}
\end{figure}

As discussed in the main text, temperature is important in the experiment: since the induced gap is closing due to the applied in-plane field, the effect of thermal quasiparticles in the response cannot be neglected. Additionally, the Al, as parent superconductor, is also affected for large enough temperatures. We measured the temperature dependence for all three resonators at different in-plane fields amplitudes and angles. The results are shown in Fig.~\ref{fig:Tdep}. The reference curve (black) is taken for each resonator without applied magnetic field. For temperatures up to 400\ mK, the frequency shifts very little and the quality factor drops. For larger temperatures, thermal quasiparticles are created which increase losses further and decrease the superfluid density, changing strongly the resonance frequency \cite{mattis1958theory,Turneaure1991_SurfaceImpedance,tinkham2004introduction}.  

For small field, $|B|=$30\ mT (red), the behavior is similar to the reference for both angles of the applied in-plane field independently of the relative orientation. We can appreciate comparing red full and empty symbols that the resonance shift is stronger when the field in perpendicular to the current in the wire (see (a) and (e)). The quality factor is sensitive to this difference as well (see (b) and (f)). This small field corresponds still to a situation where the BFS's impact is not dominant.

The situation is more interesting for $|B|=$60\ mT (blue). Looking at Fig.~2 in the main text, this value correspond to an intermediate situation where the BFS are already dominant, though not yet fully developed. There is a big difference between the field parallel and perpendicular to the wire. In the $0^\circ$-resonator in (a), we observe a non-monotonic change of the resonance between 200\ mK and 400\ mK when the field is perpendicular to the wire (blue empty symbols). This increase of $f_\mathrm{r}$ is also accompanied by a slight increase in the quality factor. A similar feature is observed for the $45^\circ$-resonator in (b). The quality factor stabilizes in this temperature window. As discussed in the theory section below, this non-monotonic features could in principle be expected from the smoothing of the change in the superfluid stiffness once the BFS open (see Fig.~\ref{fig:ns_shfit-2}).

\clearpage
\twocolumngrid
\section{Theory}
\label{ap:model}

The non-monotonic dependence of the stiffness observed in the experiment motivates the study of a possible transition to a qualitatively different state. As we show below, by considering different possible scenarios, this phase is ultimately stabilized by finite momentum superconductivity. The driving mechanisms that lead to this state are orbital effects (a Doppler/Meissner mechanism) associated with the penetration of the magnetic field into the interfacial region between the 2DEG and the bulk superconductor, and Zeeman spin-splitting combined with spin-orbit coupling in the 2DEG. Our aim is to show how the finite-momentum superconducting state corresponds to the new minimum of the free energy.

\begin{figure}[t!] 	
	\centering
\includegraphics[width=0.5\textwidth]{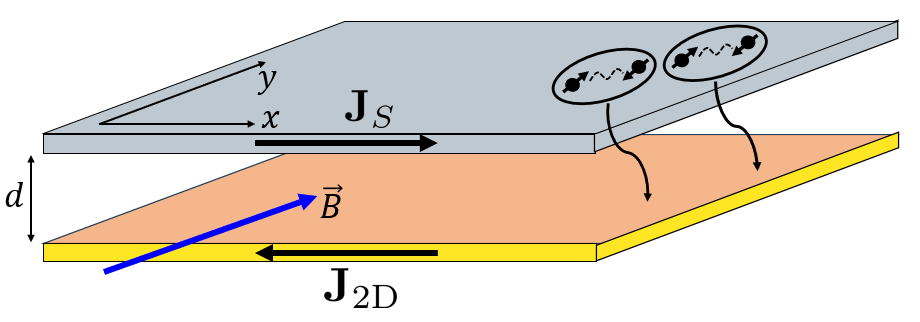} 
    \caption{Schematic representation of the two 2D layers}
    \label{fig:theo_model}
\end{figure}

To describe this effect correctly, we consider a two-layer model (see Fig.~\ref{fig:theo_model}) where the parent superconductor (the Al layer) is separated from the 2DEG layer by a distance $d$. The response of the system to the penetration of the in-plane field is to induce screening currents in both layers, which can be associated with finite orbital momenta. The finite-momentum pairing in the Al is also inherited by the 2DEG due to the coupling of the layers (proximity effect). In our microwave experiment, the value of this finite momentum is determined by a minimization of the free energy, which guarantees a zero net ground-state current for the full system: 2DEG + Al. Regardless of the mechanism, when Bogoliubov Fermi Surfaces (BFS) emerge, this minimum is always non-zero.

\subsection*{Model for the 2DEG proximitized by the superconducting Al}
We describe the InAs 2DEG system proximitized by the superconducting Al as two coupled systems (see Fig.~\ref{fig:theo_model}).

The 2DEG in the $x,y$ plane is described in the basis $\boldsymbol{\psi}_{1}(\boldsymbol{r})=\left(\psi_{1,\uparrow}(\boldsymbol{r}),\psi_{1,\downarrow}(\boldsymbol{r})\right)^T$ by the Hamiltonian
\begin{equation}
{\cal H}_{\rm 2D}=\int d^2\boldsymbol{r} \; \boldsymbol{\psi}^\dagger_{1}(\boldsymbol{r})\left[  \frac{1}{2m_1}\hat{\boldsymbol{p}}^2 -\mu+  \hat{H}_{\rm SOC}+\hat{H}_{\rm Z}\right]\;\boldsymbol{\psi}_{1}(\boldsymbol{r}),
\end{equation}
where the first term is the kinetic dispersion relation, $\mu$ the chemical potential, and the other two are the Rashba spin-orbit coupling (SOC) and the Zeeman Hamiltonians,
\begin{equation}\label{hcont1}
    \hat{H}_{\rm SOC}=-  {\boldsymbol n}_z \cdot \left({\boldsymbol \sigma} \times \hat{\boldsymbol \lambda}(\boldsymbol{r})\right),\;\;\;\;\;\;\;\;\; \hat{H}_{\rm Z}=-V_Z {\boldsymbol n}_B \cdot {\boldsymbol \sigma}.
\end{equation}
Here the magnetic field is applied in-plane in the direction $\boldsymbol{n}_B=  \cos \theta_B\hat{\boldsymbol{x}}+\sin \theta_B \hat{\boldsymbol{y}}$ given by the angle $\theta_B$ and the Zeeman field is $V_Z=\frac{1}{2}g \mu_B B $, being $g$ the g-factor and $B$ the strength of the magnetic field, while ${\boldsymbol \sigma}$ is the vector of Pauli matrices. We considered a continuum model for electrons in 2D with effective mass $m_1$ and Rashba SOC, ${\boldsymbol \lambda}(\boldsymbol{r})=\alpha_R/\hbar \left(\hat{p}_x,\hat{p}_y, 0 \right)$.

The Hamiltonian describing the Al layer in the basis $\boldsymbol{\psi}_{2}(\boldsymbol{r})=\left(\psi_{2,\uparrow}(\boldsymbol{r}),\psi_{2,\downarrow}(\boldsymbol{r})\right)^T$ is,
\begin{align}
    {\cal H}_{\rm Al}&=\int d^2\boldsymbol{r} \; \left\{\boldsymbol{\psi}^\dagger_{2}(\boldsymbol{r})\left[  \frac{1}{2m_2}\hat{\boldsymbol{p}}^2 -\mu\right]\;\boldsymbol{\psi}_{2}(\boldsymbol{r}) \right. \nonumber \\
    &-\left. \left[\Delta_S e^{i 2\boldsymbol{q} \cdot \boldsymbol{r} }{\psi}^\dagger_{2,\uparrow}(\boldsymbol{r}){\psi}^\dagger_{2,\downarrow}(\boldsymbol{r}) + \rm{h.c.}\right]
    \right\},
\end{align}
where we include a pairing with momentum $2\boldsymbol{q}$. For simplicity we neglect the Zeeman term because of the small g-factor of the Al.

The coupling between the two layers includes the penetration of the magnetic field through a Peierls' phase,
\begin{equation}\label{hcont}
    {\cal H}_c= t_c\sum_{\sigma} \int d^2\boldsymbol{r} \left( e^{i \frac{e}{\hbar}\boldsymbol{A} \cdot \boldsymbol{r}}
    \psi^\dagger_{1,\sigma}(\boldsymbol{r}) \psi^{}_{2,\sigma}(\boldsymbol{r})+ {\rm h.c. } \right),
\end{equation}
being $\boldsymbol{A}=d \left(B_y \hat{\boldsymbol{x}}- B_x \hat{\boldsymbol{y}} \right)$. Here, we assume that the magnetic field penetrates in a region of thickness $d$, defining the interface between the 2DEG and the bulk superconducting Al. 

The orbital effect generated by the magnetic field in the 2DEG is accounted by the ``Doppler shift''
\begin{equation}\label{qb}
{\boldsymbol q}_D= e \boldsymbol{A}/\hbar= d/\ell_B^2\left(\sin \theta_B  \hat{\boldsymbol{x}}- \cos \theta_B \hat{\boldsymbol{y}}\right)= d/\ell_B^2 \; \hat{\boldsymbol{B}} \times \hat{\boldsymbol z},
\end{equation}
being $\ell_B=\sqrt{\hbar/eB}$  the magnetic length.

We implement a gauge transformation $\psi_{1,\sigma}(\boldsymbol{r})\rightarrow \psi_{1,\sigma}(\boldsymbol{r}) e^{i {\boldsymbol q}_D \cdot \boldsymbol{r}}$, and we
expand the field operators as follows,
\begin{eqnarray}
     \boldsymbol{\psi}_{j}(\boldsymbol{r})&=& \frac{1}{\sqrt{\cal V}} \sum_{\boldsymbol k} e^{i {\boldsymbol k} \cdot \boldsymbol{r}} \boldsymbol{c}_{{\boldsymbol k},j}, \;\;\;\;\; j=1,2.
\end{eqnarray}
with $\boldsymbol{c}_{{\boldsymbol k},j}=\left(c_{\boldsymbol{k},j,\uparrow},c_{\boldsymbol{k},j,\downarrow}\right)^T$. Hence, the coupling Hamiltonian is written as
\begin{equation}\label{hcont1}
    {\cal H}_c=t_c \sum_{{\boldsymbol k},\sigma} \left( c^\dagger_{\boldsymbol{k},1,\sigma} c^{}_{\boldsymbol{k},2,\sigma}+{\rm h.c.}\right),
\end{equation}
while the Hamiltonian for the Al layer in the $\boldsymbol{k}$-basis is
\begin{eqnarray}
     {\cal H}_{\rm Al} &=& \sum_{\boldsymbol{k}} \; \left\{\xi_2(\boldsymbol{k}) \boldsymbol{c}^\dagger_{\boldsymbol{k},2} \boldsymbol{c}^{}_{\boldsymbol{k},2} \right.  \nonumber\\
    & & \left. -\Delta_S\left[ {c}^\dagger_{\boldsymbol{k}+\boldsymbol{q},2,\uparrow} {c}^\dagger_{-\boldsymbol{k}+\boldsymbol{q},2,\downarrow} + {\rm h.c. }\right]\right\}.
\end{eqnarray}

In order to simplify the model, we describe the effect of the Al on the 2D system in second order of perturbation theory in the coupling $t_c$. We consider the Green's function for the 2D system and include the coupling to the superconductor in terms of a self-energy. In the Nambu basis, using the matrices $\tau^{x,y,z}$ acting on the particle-hole degrees of freedom, the local self-energy reads
\begin{equation}\label{sigma}
    \Sigma(\varepsilon)= t_c^2 \tau^z G_S(\varepsilon) \tau^z,
\end{equation}
where 
\begin{equation}
   G_S(\varepsilon) 
   \simeq -\pi \nu_{\rm Al}\frac{\varepsilon+\Delta_S\tau^x}{\sqrt{\Delta_S^2-\varepsilon^2}},
\end{equation}
being $\nu_{\rm Al}$ the density of states of the Al. Substituting in Eq. (\ref{sigma}) and focusing on $\varepsilon\sim 0$, the real part adds to  the Hamiltonian ${\cal H}_{2D}({\boldsymbol k})$ the pairing term
\begin{equation}\label{pairing0}
{\cal H}_{\Delta}(\boldsymbol{ k})=  -\Delta_{0}   c^{\dagger}_{\boldsymbol{k}+\boldsymbol{q},1,\uparrow}
c^{\dagger}_{-\boldsymbol{k}+\boldsymbol{q},1,  \downarrow}+ {\rm h. c.},
\end{equation}
with $|\Delta_0|\simeq \pi t_c^2 \nu_{\rm Al} $.

Hence, the Hamiltonian of the 2DEG in $\boldsymbol{k}$-space 
is given by
\begin{align}
     {\cal H}_{\rm 2D}&=\sum_{\boldsymbol{k}} \boldsymbol{c}^\dagger_{\boldsymbol{k},1} \left[\xi_1(\boldsymbol{k}+{\boldsymbol q}_D) +  \hat{H}_{\rm Z} +\hat{H}_{\rm SOC}(\boldsymbol{k}+{\boldsymbol q}_D) \right]      \boldsymbol{c}^{}_{\boldsymbol{k},1} \nonumber \\ 
     &- \Delta_{0} \sum_{\boldsymbol{k}}  \left(c^{\dagger}_{\boldsymbol{k}+\boldsymbol{q},1,\uparrow}c^{\dagger}_{-\boldsymbol{k}+\boldsymbol{q},1,\downarrow}+ {\rm h. c.}\right). 
\end{align}
In both the 2DEG and Al Hamiltonians, the kinetic energy referred to the chemical potential is $\xi_\alpha(\boldsymbol{k})=\hbar^2 k^2/2m_\alpha-\mu$, with $k=|\boldsymbol{k}|$.
The Doppler shift contribution is incorporated in the normal terms, while the induced pairing acts on finite-momentum partners.

The resulting Bogoliubov-de-Gennes (BdG) Hamiltonian in the Nambu basis $\Psi_{\boldsymbol{\boldsymbol k}}=\left(c^{}_{\boldsymbol{\boldsymbol k}+\boldsymbol{q},1,\uparrow},c^{}_{\boldsymbol{\boldsymbol k}+\boldsymbol{q},1,\downarrow},
c^\dagger_{-\boldsymbol{\boldsymbol k}+\boldsymbol{q},1,\downarrow},-c^\dagger_{-\boldsymbol{\boldsymbol k}+\boldsymbol{q},1,\uparrow}\right)^T$, reads
\begin{align}\label{bdg-cont}
    H^{\rm BdG}_{\rm 2D}&(\boldsymbol{k},\boldsymbol{q}) =
    \tau^+\;\left[\xi_1({\boldsymbol{k}+{\boldsymbol q}_D}+\boldsymbol{ q})\sigma^0+H_{\rm SOC}(\boldsymbol{k}+{\boldsymbol q}_D+\boldsymbol{q})\right]  \nonumber\\ 
    &-
    \tau^-\;\left[\xi_1({-\boldsymbol{k}+{\boldsymbol q}_D}+\boldsymbol{ q})\sigma^0+H_{\rm SOC}(-\boldsymbol{k}+{\boldsymbol q}_D+\boldsymbol{q})\right]  \nonumber \\
   &+ 
   \tau^0 H_{\rm Z} - \Delta_0\; \tau^x \sigma^0,
\end{align}
where $\tau^\pm=(\tau^0\pm\tau^z)/2$. Notice that, after a gauge transformation, this Hamiltonian
can be also expressed  as one with a pairing term with a finite momentum $2(\boldsymbol{q}+ {\boldsymbol q}_D)$.

Respectively, the BdG Hamiltonian for the Al layer in the Nambu basis $\Psi_{\boldsymbol{\boldsymbol k},2}=\left(c^{}_{\boldsymbol{\boldsymbol k}+\boldsymbol{q},2,\uparrow},c^{}_{\boldsymbol{\boldsymbol k}+\boldsymbol{q},2,\downarrow},
c^\dagger_{-\boldsymbol{\boldsymbol k}+\boldsymbol{q},2,\downarrow},-c^\dagger_{-\boldsymbol{\boldsymbol k}+\boldsymbol{q},2,\uparrow}\right)^T$, reads
\begin{align}\label{bdg-cont-Al}
    H^{\rm BdG}_{\rm Al}&(\boldsymbol{k},\boldsymbol{q}) =
    \tau^+\;\left[\xi_2({\boldsymbol{k}}+\boldsymbol{ q})\sigma^0\right]  -
    \tau^-\;\left[\xi_2({-\boldsymbol{k}}+\boldsymbol{ q})\sigma^0\right]  \nonumber \\
   & - \Delta_S\; \tau^x \sigma^0.
\end{align}

An important aspect to highlight is the fact that ${\boldsymbol q}_D$ is an external parameter defined by the value of the magnetic field as indicated in 
Eq. (\ref{qb}). Instead, the finite momentum $\boldsymbol{q}$ is a {\em variational parameter}, which is determined by the requirement that the full system: 2D + Al achieves a state of minimum free-energy.

\subsection*{Superfluid stiffness}
The simplest way to define the superfluid stiffness $D$ is as a measure of the energy cost of phase fluctuations from the equilibrium mean-field configuration of the superconductor \cite{schmidt2013physics,tinkham2004introduction}. To directly bridge this macroscopic thermodynamic quantity with our microscopic Bogoliubov-de Gennes formalism, it is convenient to characterize the order parameter in terms of the single-electron phase $\phi(\boldsymbol{r})$ such that
$\Delta(\boldsymbol{r})=|\Delta(\boldsymbol{r})| e^{i 2 \phi(\boldsymbol{r})}$. The increase in free energy due to a phase gradient is then  
\begin{equation}\label{df}
    \Delta F=2\int d\boldsymbol{r} \sum_{j,l} D_{jl} \partial_j \phi(\boldsymbol{r}) \partial_l \phi(\boldsymbol{r}).
\end{equation}
The stiffness is connected to the London Kernel, i.e. the supercurrent response to an external perturbation
$\boldsymbol{A}^{\rm ext}$, 
\begin{equation}
    J_j=-\left(\frac{2e}{\hbar}\right)^2 \sum_{l} D_{jl} A^{\rm ext}_l,
\end{equation}
being  $[\lambda^{-2}]_{jl}=\mu_0(2e/\hbar)^2 D_{jl}$ the London penetration depth, with $\mu_0$ the vacuum permeability.

A practical way to calculate it is by introducing twisted boundary conditions in one of the directions \cite{Scalapino1993Apr}. In our case, we consider a 2D system of dimension $L\times W$  and fix the direction of the current in the inductor of the lumped-element resonators along $x$ and impose that the wave function satisfies $\Psi(x+L)=e^{i \theta} \Psi(x)$. This is a gauge-equivalent problem such that the 2D system is bent to form a torus with an Aharonov-Bohm flux $\Phi$ inside it, so that the associated vector potential is $\boldsymbol{A}^{\rm ext}=(\Phi/L)\hat{\boldsymbol{x}}$ and $\theta=-\frac{e}{\hbar}\Phi$. We define $\boldsymbol{\phi}= -(\hbar/e)\boldsymbol{A}^{\rm ext}$, such that the kinetic term in the minimal coupling $\boldsymbol{p} \rightarrow \boldsymbol{p}-e\boldsymbol{A}^{\rm ext}$, becomes $\boldsymbol{k} \rightarrow \boldsymbol{k}+\boldsymbol{\phi}$.

Here we are explicitly distinguishing the vector potential associated with the in-plane magnetic field $\boldsymbol{A}$ which brings the finite momentum $\boldsymbol{q}$, from $\boldsymbol{A}^{\rm ext}$ which is the microwave perturbation field that probes the system and is considered through $\boldsymbol{\phi}$. Formally, the free-energy will be a function of $\boldsymbol{q}$ and $\boldsymbol{\phi}$, $F(\boldsymbol{q}+\boldsymbol{\phi})$. Our aim is to analyze $F(\boldsymbol{q},\boldsymbol{\phi}=0)$ to find the value ${\boldsymbol q}_0$ at which it is a minimum and evaluate the stiffness at this momentum. Therefore, we redefine ${\boldsymbol q} \equiv \boldsymbol{q}+\boldsymbol{\phi}$, and we calculate the
stiffness directly as 
\begin{equation}\label{t0}
D_{jl}=\frac{1}{4}\left.\frac{\partial^2 F({\boldsymbol q})}{\partial q_j \partial q_l}\right|_{{\boldsymbol q}_0}.
\end{equation} 

The free energy  of the 2DEG can be expressed in terms of the eigenenergies $ E_{\boldsymbol{ k},j}({\boldsymbol q})$ of the Hamiltonian of Eq. (\ref{bdg-cont})
\begin{align}\label{free}
    &F^{\rm (2D)}({\boldsymbol q})=LW \int_{|\boldsymbol{k}|<\Lambda} \frac{d^2\boldsymbol{k}}{(2\pi)^2} \\
    &\;\;\;\;\;\left\{-\frac{1}{2 \beta}\sum_j \ln \left(1+e^{-\beta E_{\boldsymbol{k},j}(\Delta_0(T),{\boldsymbol q}) }\right)
     +\xi_1({\boldsymbol{k}+{\boldsymbol q}})\right\},\nonumber
\end{align}
being $\beta=1/(k_BT)$ for temperature $T$. Here, we assume that the induced pairing potential $\Delta_0(T)$ follows the standard BCS temperature dependence, which we approximate using $\Delta_0(T) \approx \Delta_0(0) \tanh\left(1.74 \sqrt{T_c/T - 1}\right)$ as is standard for extracting kinetic inductance from microwave resonators \cite{annunziata2010tunableinductors,tinkham2004introduction}.
We also have a similar expression for the Al.

We have verified that the results are equivalent to calculating the diamagnetic component of the current-current response function, which is the alternative procedure to calculate the superfluid stiffness \cite{Babkin2024BFS}. We find that there is more information regarding the interpretation of the change in the stiffness using this thermodynamic approach. We continue the discussion further below.

\begin{figure*}[t!] 	
	\centering
\includegraphics[width=0.9\textwidth]{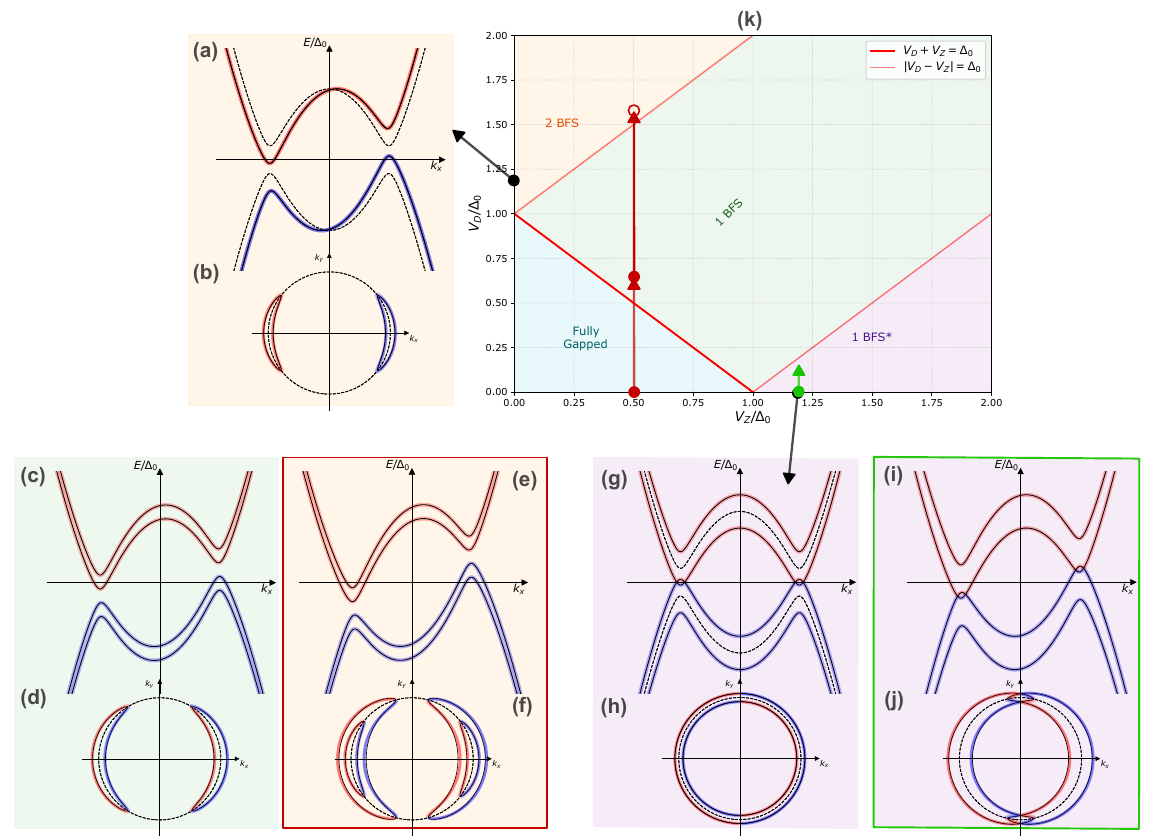} 
    \caption{\textbf{Bogoliubov Fermi Surfaces without SOC:} \textbf{(a,c,e,g,i)} Energy bands and \textbf{(b,d,f,h,j)} Fermi surface contours obtained for the BdG Hamiltonian in our simplified model of the proximitized 2DEG. Dashed lines correspond to zero field. Colours correspond to the numerical solutions and black lines to the analytic expressions (see text). \textbf{(a,b)} Doppler energy $V_D>\Delta_0$, Zeeman energy $V_Z=0$. \textbf{(a)} electron-like (red) and hole-like (blue) bands are tilted with applied field (here in the $y$-direction) crossing the Fermi level. \textbf{(b)} As a consequence ``banana-shaped'' BFS emerge along the $k$-direction perpendicular to the field. \textbf{(g,h)} Doppler energy $V_D=0$, while Zeeman energy $V_Z>\Delta_0$. \textbf{(g)} bands are shifted with applied field (here in the $y$-direction) crossing the Fermi level. \textbf{(h)} ``circular'' BFS emerge in the $k$-space. There is no anisotropy. \textbf{(c-f)} A combination of both effects ($V_Z,V_D$) results in the successive transitions: gapped phase $\rightarrow$ \textbf{(c,d)} 1-BFS $\rightarrow$ \textbf{(e,f)} 2-BFS as summarized in the phase diagram \textbf{(k)}. \textbf{(c,d)} $V_Z<\Delta_0$, $V_D<\Delta_0$ but $V_Z+V_D >\Delta_0$. The 1-BFS pair is formed corresponding to the band-crossing for a given spin. \textbf{(e,f)} $V_D-V_Z >\Delta_0$. The 2-BFS phase emerge. \textbf{(i-j)} $|V_D-V_Z|>\Delta_0$. There are 2-BFS which evolve from the ``circular-shape'' into ``banana-shape'' as discussed in the text. }
    \label{fig:supple1}
\end{figure*}

\subsection*{Spectral properties and ground state energy for special cases}
We now focus on the analysis at $T=0$, where some analytical results can be derived. In this case, the free-energy Eq. (\ref{free}) coincides with the ground-state energy (GSE) 
\begin{eqnarray} \label{inas}
    E^{\text{2D}}_{\rm GS}({\boldsymbol q})&=& LW \int_{|\boldsymbol{k}|<\Lambda} \frac{d^2\boldsymbol{k}}{(2\pi)^2} {\cal E}_{\boldsymbol{k}}({\boldsymbol q}),
\end{eqnarray}
where
\begin{eqnarray*} 
    {\cal E}_{\boldsymbol{k}}({\boldsymbol q})&=& \sum_{j}  \theta\left(-E_{\boldsymbol{k},j}({\boldsymbol q})\right) E_{\boldsymbol{k},j}({\boldsymbol q})/2 + 
     \xi_{\boldsymbol{k}+{\boldsymbol q}}.
\end{eqnarray*}
Here, the sum over $j$ runs over the negative eigenenergies, i.e. satisfying $ E_{\boldsymbol{ k},j}({\boldsymbol q})<0$.

To calculate the GSE of the full system: 2DEG+Al, we must add the contribution of the Al to Eq. (\ref{inas}). At leading order in $\boldsymbol{q}$ this
component 
can be expressed as follows,
\begin{equation}\label{e0-al}
    E^{\text{Al}}_{\rm GS}({\boldsymbol q})=E_{\rm GS}^{\rm BCS, Al}+2D^{\text{Al}}|{\boldsymbol q}|^2,
\end{equation}
where
\begin{equation}
E_{\rm GS}^{\rm BCS, Al}= LW \int_{|\boldsymbol{k}|<\Lambda} \frac{d^2\boldsymbol{k}}{(2\pi)^2} \left[- \sqrt{\xi_2^2({\boldsymbol{k}})+\Delta_S^2}+\xi_2({\boldsymbol k})\right],
\end{equation}
is the GSE of a 2D BCS model with zero-momentum pairing, without Zeeman field, while the other term is the correction introduced by the finite-momentum of the Cooper pairs.

\subsubsection*{No Zeeman, no SOC but with Doppler shift}

In what follows, we simplify the notation by omitting the subindex in the kinetic energy of the 2DEG: $\xi_1(\boldsymbol{k}) \rightarrow \xi_{\boldsymbol{k}}$. In this case, Eq. (\ref{bdg-cont}) reduces to 
\begin{equation}\label{bdg-cont-2}
    H_{\rm BdG}(\boldsymbol{k},{\boldsymbol q}') = \xi_{\boldsymbol{k}+{\boldsymbol q}'} \tau^+\sigma^0 - \xi_{-\boldsymbol{k}+{\boldsymbol q}'} \tau^-\sigma^0 -\Delta_0 \; \tau^x\sigma^0,
\end{equation}
where ${\boldsymbol q}'={\boldsymbol q}_D+{\boldsymbol q}$. The spectrum consists of two degenerate eigenenergies for each $\boldsymbol{k}$,
\begin{align}\label{dis-2-1}
    E_{\boldsymbol{k},\pm}&= \frac{(\xi_{\boldsymbol{k}+{\boldsymbol q}'}-\xi_{-\boldsymbol{k}+{\boldsymbol q}'})}{2}\nonumber \\
    &\pm \frac{1}{2} \sqrt{(\xi_{\boldsymbol{k}+{\boldsymbol q}'}+\xi_{-\boldsymbol{k}+{\boldsymbol q}'})^2+
    4 \Delta_0^2}\nonumber \\
    & \simeq \;  \hbar \boldsymbol{v}_F  \cdot {\boldsymbol q}' \pm \sqrt{\left(\xi_{\boldsymbol{k}}+\kappa \;{\boldsymbol q}'^2\right)^2+ \Delta_0^2},
\end{align}
being $\kappa= \hbar^2/2m$. We see clearly the effect of the Doppler shift which is to tilt the bands. The superconducting gap closes in $\boldsymbol{k}$ points satisfying $\Delta_0=\hbar\; \boldsymbol{
v}_F \cdot {\boldsymbol q}'$ 
with the development of  spin-degenerate Bogoliubov Fermi surfaces. 

In Fig.~\ref{fig:supple1} we compare the numerical solutions with the analytic expressions developed in this section. In (a) we show the bands in the presence of Doppler and compare with the Eq.~(\ref{dis-2-1}). In what follows, we focus on ${\boldsymbol q}' || \hat{\boldsymbol{x}}$, perpendicular to the direction of the applied magnetic filed. In order to get analytical expressions, we also focus on $|{\boldsymbol q}|'$ close to the critical value at which the BFS emerge. Under this assumption, we neglect the term $\kappa {\boldsymbol q}'^2$ in Eq. (\ref{dis-2-1}). The BFS formed by the lower band crossing zero energy from negative energies  
can be described in polar coordinates $(k,\varphi)$ as follows
\begin{equation}\label{pock-dop}
    k_{\pm} \simeq \frac{1}{\sqrt{\kappa}} \sqrt{\mu\pm \sqrt{E_{{\boldsymbol q}'}^2 \cos^2\varphi-\Delta_0^2}},\;\;\;\; |\varphi|<\varphi_c,
\end{equation}
being $E_{{\boldsymbol q}'}=\hbar v_F |{\boldsymbol q}'|$, with $E_{{\boldsymbol q}'}\geq \Delta_0$, and
$\varphi_c=\arccos(\Delta_0/E_{{\boldsymbol q}'})$. We see that the minimal momentum ${\boldsymbol q}'$ required
for the development of the BFS is
\begin{equation}
q_c=\frac{\Delta_0}{\hbar v_F}.
\end{equation}
The BFS corresponding to the crossing of the positive energies are defined by  the same $k$-values, but the angle runs 
within $|\varphi-\pi|<\varphi_c$. In Fig.~\ref{fig:supple1}(b) we show the contours obtained with the expressions from Eq.~(\ref{pock-dop}) which fit very well the numerical result. 

The result for the GSE when the BFS emerge is
\begin{equation}\label{e0-qb}
E_{\rm GS}^{\rm (2D)}({\boldsymbol q}')=E_{\rm GS}^{\rm (0)}({\boldsymbol q}')+ \theta(|\boldsymbol{ q}'|-q_c) E_{\rm GS}^{\rm BFS}({\boldsymbol q}'),
\end{equation}
where the first term is the integral without taking into account the BFS and it is given by
\begin{eqnarray}\label{e0-wp}
    E_{\rm GS}^{\rm (0)}({\boldsymbol q}') &=&  LW \int_{|\boldsymbol{k}|<\Lambda} \frac{d^2\boldsymbol{k}}{(2\pi)^2} \;\left\{
    \xi_{\boldsymbol{k}+{\boldsymbol q}'}\right.\nonumber\\
   & & \left. \;\;\;\;\; \;\;\;\;\;-\sqrt{(\xi_{\boldsymbol{k}}+\kappa {\boldsymbol q}'^2)^2 +  \Delta_0^2} \right\}.
\end{eqnarray}
The leading contribution of this term is
\begin{eqnarray}\label{wp}
E_{\rm GS}^{(\rm 0)}({\boldsymbol q}')&\simeq &E_{\rm GS}^{\rm (2D,BCS)}(0) +\beta|{\boldsymbol q}'|^2, \nonumber\\
\beta&=& (\hbar v_F)^2 {\cal N}_0,
\end{eqnarray}
with ${\cal N}_0=LW/(2 \pi \kappa)$.

The other term in Eq. (\ref{e0-qb}) takes into account the change in energy originated by the formation of the BFS and it reads
\begin{eqnarray}\label{e0-bfs}
&  E_{S}^{\rm BFS}({\boldsymbol q}') =
2 L W \int_{S} \frac{d^2\boldsymbol{k}}{(2\pi)^2} \;  E_{\boldsymbol{k},+}({\boldsymbol q}'),
\end{eqnarray}
with $E_{\boldsymbol{k},+}({\boldsymbol q}')$ given by Eq. (\ref{dis-2-1}).
The region $S$ of the $\boldsymbol{k}$-plane is enclosed by the BFS formed by the crossing of the band with positive energy. This is defined by $k$ values
between $k_+(\varphi)$ and $k_-(\varphi)$ given by Eq. (\ref{pock-dop}) and angles $|\varphi-\pi|<\varphi_c$.  
We can calculate approximately this energy for $|{\boldsymbol q}'|\gtrsim q_c$. The result is
\begin{eqnarray}\label{DS}
 E_{\rm GS}^{\rm BFS}({\boldsymbol q}')  &\simeq& L W  \int_{-\varphi_c}^{\varphi_c} \frac{d\varphi}{(2\pi)^2}
\left\{
\left[k_+^2(\varphi)-k_-^2(\varphi)\right] \right. \nonumber\\
& & \left.\;\;\;\;\;\;\;\;\;\;\;\; \times\left[\Delta_0-E_{{\boldsymbol q}'}\cos(\varphi)\right]\right\}
\nonumber\\
& & \;\;\;\;\;\;\; \simeq -\frac{40 {\cal N}_0}{6\pi}(\hbar v_F)^2(q\lt{'}-q_c)^2,
\end{eqnarray}
where in the last step we have approximated $\cos(\varphi) \sim 1-\varphi^2/2$. 

Adding the different contributions  leads to the following behavior of the GSE for the joint 2D system and the superconducting bulk,
\begin{align}\label{gs-1}
    E_{\rm GS}^{\rm tot}({\boldsymbol q})&=
    E_{\rm GS}^{\rm BCS}(0)+ \theta(|{\boldsymbol q}+{\boldsymbol q}_D|-q_c)E^{\text{BFS}}_S(\boldsymbol{q}+\boldsymbol{q}_D)\nonumber\\
    &+\beta|{\boldsymbol q}+{\boldsymbol q}_D|^2+2D^{\rm Al} |{\boldsymbol q}|^2,
\end{align}
with $E_{\rm GS}^{\rm BCS}$ being the ground state energy of the 2DEG+Al, while $ \beta >0$ has been defined in Eq. (\ref{wp}). The second term describes the change of behavior due to the formation of the BFS. In Fig.~\ref{fig:supple2}(a,b) we present the result of $E_{\rm GS}^{\rm 2D}(q_x)$ calculated for increasing field considering Doppler energy but no Zeeman energy, and without SOC. The field is (a) perpendicular and (b) parallel to the current, taken along $\hat{x}$.

 \subsubsection*{Doppler shift, Zeeman but no SOC}
In this case, the spectrum is given by bands labeled by $(s,\pm)$, where $s=\pm$ is related to the spin splitting due to the Zeeman field,
\begin{align}\label{dis-1-1}
    E_{\boldsymbol{k},s,\pm}&({\boldsymbol q}')= s V_Z +  \frac{(\xi_{\boldsymbol{k}+{\boldsymbol q}'}-\xi_{-\boldsymbol{k}+{\boldsymbol q}'})}{2} \nonumber \\
    &\pm \frac{1}{2} \sqrt{(\xi_{\boldsymbol{k}+{\boldsymbol q}'}+\xi_{-\boldsymbol{k}+{\boldsymbol q}'})^2+
    4\Delta_0^2)}\nonumber \\
   &\simeq s V_Z +   \boldsymbol{v}_F \cdot {\boldsymbol q}' \pm \sqrt{[\xi_{\boldsymbol{k}}+\kappa ({\boldsymbol q}')^2]^2+ \Delta_0^2}.
\end{align}

Depending on the relative magnitude of $V_Z$ and the Doppler energy $V_D=\hbar v_F q_D$, two scenarios may take place, which lead to the emergence of a single pair or a double pair of BFS (see phase diagram in Fig.~\ref{fig:supple1}(k)). We discuss them in two steps. 

The first pair of BFS emerge when the bands  $(-,+)$ and $(+,-)$ cross zero energy (see Figs.~\ref{fig:supple1}(c,g)). For ${\boldsymbol q}' =0$ (Fig.~\ref{fig:supple1}(g)), the superconducting gap closes when $V_Z=\Delta_0$ and, for $V_Z>\Delta_0$, the BFS form one or two circles, depending on the value of the chemical potential $\mu$, relative to the amplitude of the Zeeman field $V_Z$ \cite{Ruiz2025Dec}. We focus here on $\mu>V_Z$, which corresponds to two circles (see Fig.~\ref{fig:supple1}(h)). The effect of ${\boldsymbol q}'$ in this picture is to tilt the spectrum generating a splitting of these circular BFS (see  Fig.~\ref{fig:supple1}(i,j)).
 
The $\boldsymbol{k}$-values enclosing the surface formed by the crossing from negative energies reads
\begin{align}
    k^{(1)}_{\pm}\simeq &\frac{1}{\sqrt{\kappa}} \sqrt{\mu\pm \sqrt{(V_Z+E_{{\boldsymbol q}'} \cos\varphi)^2-\Delta_0^2}},\nonumber \\&|\varphi|<\varphi^{(1)}_c, \;\;\;\;   \;\;\;\;E_{{\boldsymbol q}'}\cos\varphi+V_Z \geq \Delta_0,
\end{align}
being  
\begin{eqnarray}
\varphi_c^{(1)}&=& 2 \pi, \;\;\;\;\;\;\;\;\;\;\;\;\;\;\;\;\;\;\;\;\;\;\;\;\;\;\;\;\;\;\;\;\;\;\;\; |{\boldsymbol q}'|<q_c^{(1)}, \nonumber \\
{\varphi}^{(1)}_c&=&\arccos(|\Delta_0- V_Z|/E_{{\boldsymbol q}'}),  \;\;\;\;\;\; |{\boldsymbol q}'|>q_c^{(1)},
\end{eqnarray}
with 
\begin{equation}
q^{(1)}_c=\frac{|V_Z-\Delta_0|}{\hbar v_F}.
\end{equation}

The BFS due to the crossing from positive energies is defined by the same $k$-values and angles $\varphi+\pi$. The shape of these BFS change from non-centered circles to banana-shape when $|{\boldsymbol q}'|>q^{(1)}_c$. These solutions are plot with black lines in Fig.~\ref{fig:supple1}(d,h,j).

The integration over $\boldsymbol{k}$ leads to the following expression for the GSE
\begin{equation}\label{e0-qb-z}
E_{\rm GS}^{\rm (2D)}({\boldsymbol q}')=E_{\rm GS}^{\rm (0)}({\boldsymbol q}')+ {E}_{\rm GS}^{\rm BFS}({\boldsymbol q}'),
\end{equation}
which has exactly the same structure as Eq. (\ref{e0-qb}). The first term is also given by Eq. (\ref{e0-wp}) and the second one can be also expressed by Eq. 
(\ref{e0-bfs}). In the present case, the latter contribution can be written as
\begin{eqnarray}
{E}_{\rm GS}^{\rm BFS}({\boldsymbol q}') &=& \frac{1}{2}E_{S_1}^{\text{BFS}}(\boldsymbol{q}')\theta(|{\boldsymbol q}'|-q_c^{(1)}) \nonumber \\
& &-\theta(V_Z-\Delta_0)2 V \frac{LW}{(2\pi)^2}S_1,
\end{eqnarray}
where $S_1$ is the area enclosed by the BFS. The effect of the  Doppler-shift is accounted for the first term of this equation. The factor $1/2$ is due to the fact that, in the present case, only a single spin species contributes while in Eq. (\ref{e0-bfs}) there was spin-degeneracy. The effect of the Zeeman field is accounted for the second term. 

\begin{figure}[t!]
\includegraphics[width=0.5\textwidth]{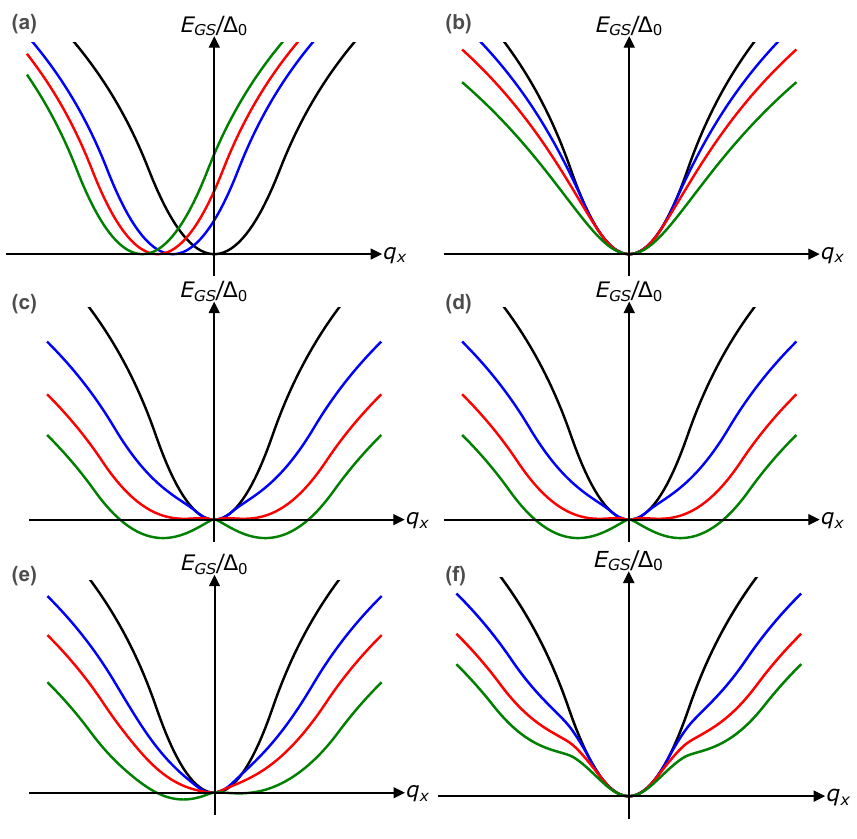} 
    \caption{\textbf{Ground-state energy vs $q_x$ in different scenarios:} In \textbf{(a-d)} there is no SOC, such that the bands correspond to Fig.~\ref{fig:supple1}. In \textbf{(e-f)} we consider also the effect of SOC, such that the bands correspond to those presented in Fig.~3 of the main text. In the case of Doppler, the result with and without SOC is very similar. \textbf{(a,b)} Increasing the value of Doppler energy $V_D$ until $V_D>\Delta_0$ keeping $V_Z=0$. The field is \textbf{(a)} perpendicular $\mathbf{B}=B\hat{y}$ and \textbf{(b)} parallel $\mathbf{B}=B\hat{x}$ to the current taken along $\hat{x}$. \textbf{(a,b)} Increasing the value of Zeeman energy $V_Z$ until $V_Z>\Delta_0$ keeping $V_D=0$. The field is \textbf{(c)} perpendicular $\mathbf{B}=B\hat{y}$ and \textbf{(d)} parallel $\mathbf{B}=B\hat{x}$ to the current. \textbf{(e,f)} Same as \textbf{(c,d)} considering SOC.} 
    \label{fig:supple2}
\end{figure}

Interestingly, the area $S_1$ is a function of ${\boldsymbol q}$, which decreases fast when the shape of the BFS changes from a circle to a banana-shape. As a consequence, starting from ${\boldsymbol q}'=0$, the GSE decreases, achieves two local minima at $|{\boldsymbol q}|'\neq 0$ and then increases for higher values of ${\boldsymbol q}'$. In Fig.~\ref{fig:supple2}(c,d) we present the result of $E_{\rm GS}^{\rm 2D}(q_x)$ calculated for increasing field considering Zeeman energy but zero Doppler energy, and without SOC. In this case, there is no difference between the field (c) perpendicular and (d) parallel to the current, taken along $\hat{x}$, since the BFS open forming a circle (see Fig.~\ref{fig:supple1}(g,h)).

As ${\boldsymbol q}'$ increases (see the red arrow in phase diagram in Fig.~\ref{fig:supple1}(k)), a second pair of BFS emerge, which  corresponds to the bands $(+,+)$ and $(-,-)$ crossing zero energy (Fig.~\ref{fig:supple1}(e)). The $\boldsymbol{k}$-values enclosing the second surface generated by the crossing from negative energies are 
\begin{align}
    k^{(2)}_{\pm}\simeq &\frac{1}{\sqrt{\gamma}} \sqrt{\mu\pm \sqrt{(-V_Z+E_{{\boldsymbol q}'} \cos\varphi)^2-\Delta_0^2}},\nonumber \\&|\varphi|<\varphi^{(2)}_c, \;\;\;\;   \;\;\;\;E_{{\boldsymbol q}'}\cos\varphi-V_Z \geq \Delta_0,\nonumber\\
\end{align}
being  ${\varphi}^{(2)}_c=\arccos((\Delta_0+ V_Z)/E_{{\boldsymbol q}'})$ for $E_{{\boldsymbol q}'} \geq \Delta_0+ V_Z$. As before, the BFS due to the crossing from positive energies is also described by these expressions upon changing $\varphi\rightarrow \varphi +\pi$. These  BFS  emerge when $|{\boldsymbol q}'|>q^{(2)}_c$, with
\begin{equation}
 q^{(2)}_c=\frac{V_Z+\Delta_0}{\hbar v_F}.
\end{equation}
The BFS described by these Eq. are shown with the black lines in Fig.~\ref{fig:supple1}(f).
 
The integration over $\boldsymbol{k}$ leads to the following expression for the GSE,
\begin{equation}\label{e0-qb-z}
E_{\rm GS}^{\rm (2D)}({\boldsymbol q}')=E_{\rm GS}^{\rm (0)}({\boldsymbol q}')+  {E}_{\rm GS}^{\rm BFS}({\boldsymbol q}'),
\end{equation}
which has exactly the same structure as Eq. (\ref{e0-qb}). The first term is also given by Eq. (\ref{e0-wp}) and the second one can be also expressed by Eq. 
(\ref{e0-bfs}). 

Taking into account both pairs of BFS, the latter contribution can be written as
\begin{eqnarray}\label{ebs-2zq}
 {E}_{\rm GS}^{\rm BFS}({\boldsymbol q}')&=&\frac{1}{2}E^{\text{BFS}}_{S_1}(\boldsymbol{q}')\theta(|{\boldsymbol q}'|-q_c^{(1)}) \nonumber\\
& & -2 V_Z\frac{LW}{(2\pi)^2} S_1 \theta(V_Z-\Delta_0)\\
& &+ 
\left[\frac{1}{2}E^{\text{BFS}}_{S_2}(\boldsymbol{q}')+2 V_Z \frac{LW}{(2\pi)^2} S_2 \right] 
\theta(|{\boldsymbol q}'|-q_c^{(2)})\nonumber 
\end{eqnarray}
where $S_1$ and $S_2$ are, respectively,  the area enclosed by the internal and external BFS.

Taking into account also the contribution of the parent superconductor, the total GSE reads
\begin{align}\label{gs-2}
    E_{\rm GS}({ \boldsymbol q})&=
    E_{\rm GS}^{\rm BCS}+ E_{\rm GS}^{\rm BFS}({\boldsymbol q}+{\boldsymbol q}_D)
    +\beta|{\boldsymbol q}+{\boldsymbol q}_D|^2+2D^{\rm Al} |{\boldsymbol q}|^2,
\end{align}
where $E_{\rm GS}^{\rm BFS}({ \boldsymbol q}+{\boldsymbol q}_D)$ is given by Eq. (\ref{ebs-2zq}). As before, $E_{\rm GS}^{\rm BCS}$ is the ground state energy of the 2DEG+Al, and $ \beta >0$ is given by Eq. (\ref{wp}).

The phase diagram is shown in Fig.~\ref{fig:supple1}.


\subsubsection*{With Doppler shift, Zeeman and SOC}
The combination of SOC and Zeeman with a local s-wave pairing potential effectively generates pairing potential in the s-wave as well as p-wave channels \cite{Ruiz2025Dec}. For a large Fermi energy, we can neglect the p-wave contribution and we get the following simple expressions for the eigenenergies 
\begin{align}\label{todo}
    E_{\boldsymbol{k},s,\pm}&= \frac{({\xi}_{{\boldsymbol{k}+{\boldsymbol q}\lt{'}},s}-\tilde{\xi}_{{-\boldsymbol{k}+{\boldsymbol q}\lt{'}},s})}{2}\nonumber\\ &\pm 
    \frac{1}{2} \sqrt{({\xi}_{{\boldsymbol{k}+{\boldsymbol q}\lt{'}},s}+\tilde{\xi}_{{-\boldsymbol{k}+{\boldsymbol q}\lt{'}},s})^2+4 \Delta_0^2},
\end{align}
where $s=\pm$ labels the two bands associated to the spin, while 
\begin{align}
    {\xi}_{\boldsymbol{k},s}&=\xi_{\boldsymbol{k}}-s |\boldsymbol{d}_{\boldsymbol{k}}+\boldsymbol{V}|, \nonumber \\
\tilde{\xi}_{\boldsymbol{k},s}&=\xi_{\boldsymbol{k}}+s |\boldsymbol{d}_{\boldsymbol{k}}+\boldsymbol{V}|,
\end{align} where we have introduced $\boldsymbol{V}=V_Z \boldsymbol{n}_B$ and $\boldsymbol{d}_{\boldsymbol{k}}=\alpha_R(k_y,-k_x,0)$ to form the effective $\boldsymbol{k}$-dependent field $\boldsymbol{d}_{\boldsymbol{k}}+\boldsymbol{V}$, generated by the combination of the magnetic field and the spin-orbit coupling.

The BFS for ${\boldsymbol q}\lt{'}=0$ have banana shapes in the present case (see Fig.~3 in the main text) and they develop when  $V_Z=\Delta_0$. However, they are significantly anisotropic and the spectrum remains gaped in the direction parallel to the magnetic field. The resulting GSE in this case is described for finite ${\boldsymbol q}'$ as follows (see \cite{YuanFu2022})
\begin{eqnarray}\label{e0-lif}
   & &  E^{\rm 2D}_{\rm GS}({\boldsymbol q}')={\rm Eq.}(\ref{gs-2})
   \nonumber \\
    & &\;\;\;\;\;\;\;\;\;- \left(\alpha_0-\alpha_1 |{\boldsymbol q}'|^2\right){\boldsymbol q}' \cdot (\hat{n}_{B}\times \hat{\boldsymbol{z}}),
\end{eqnarray}
where the term in the second line is usually named ``Lifshitz invariant'' and it is $\propto \alpha_R/\hbar$. It reflects the anisotropy of the BFS in the fact that it contributes in the direction perpendicular to the magnetic field $(\hat{n}_{B}\times \hat{\boldsymbol{z}})$ where it has two minima at ${\boldsymbol q}'_{\pm}$. Since the contribution of the SOC is described by odd powers of $\boldsymbol{q}'$,  the two minima arising of the GSE are not degenerate. This is shown in Fig.~\ref{fig:supple2}(e,f) where we present the result of $E_{\rm GS}^{\rm 2D}(q_x)$ calculated for increasing field considering Zeeman energy with SOC but zero Doppler energy. The field is (e) perpendicular and (f) parallel to the current.

\begin{figure}[t!] 	
	\centering
\includegraphics[width=0.5\textwidth]{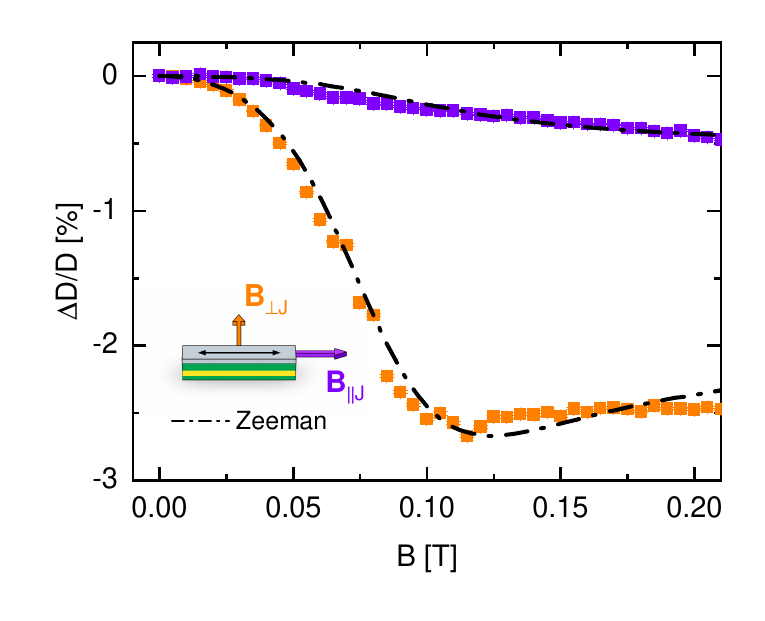} 
    \caption{Magnetic field dependence of the total stiffness for a given temperature of reference $T_{\rm eff}=200$~mK such that $\frac{k_B T_{\rm eff}}{\Delta_0}=0.215$, $\Delta_{\rm S}=0.2$~meV, $\Delta_0=0.08$~meV, 
    $\Delta_{\rm S}(T_{\rm eff})/\gamma_{\parallel}=$115992, $\gamma_{\perp}=4\gamma_{\parallel}$, $g=34.98$, $B_c=0.079$~T, $\mu=50.6$~meV, $m=0.0403m_e$, $\lambda_{SOC}=15$~meV nm.}
    \label{fig:ns_shfit-1}
\end{figure}

\section{From the theory to the experimental data}
\label{ap:fitting}

We recall that our model relies on the assumption that the Al and 2DEG are weakly interacting systems. We relate the measured frequency shift $\Delta f_r$ (see Eq. (2) of the main text) with the superfluid stiffness
\begin{equation}\label{rel-frec}
    \frac{\Delta D}{D}=
    \frac{\delta D^{\rm Al}(T)+\tilde{\gamma}\delta D^{\rm 2D}(B,T)}{D^{\rm Al}(T_{\rm eff})+ \tilde{\gamma} D^{\rm 2D}(B=0,T_{\rm eff})},
\end{equation}
where $\delta D(X)\equiv D(X)-D(X_{\rm ref})$ is the change of the stiffness at temperature and magnetic field $B$ with respect to the one at a reference temperature $T_{\rm eff}$ and magnetic field $B=0$. We assume that the stiffness of the Al layer varies only on $T$ and not on $B$, within the range of in-plane magnetic fields considered. Furthermore,  following Ref. \cite{annunziata2010tunableinductors}, the Al stiffness is assumed to be proportional to the BCS temperature dependence of the gap,
\begin{equation}
    D^{\rm Al}(T)=K \;\Delta_{\rm S}\tanh \left( \frac{\Delta_{\rm S}}{2k_B T}\right), \;\;\;\;\;\;T\ll T_c,
\end{equation}
being $K$ a constant factor in the low-temperature regime. Substituting in Eq. (\ref{rel-frec}) we get 
\begin{align*}
    \frac{\Delta D}{D}&=\\&\frac{\Delta_{\rm S}\left[\tanh \left( \frac{\Delta_{\rm S}}{2k_B T}\right)-\tanh \left( \frac{\Delta_{\rm S}}{2k_B T_{\rm eff}}\right)\right]+\gamma\delta D^{\rm 2D}(B,T)}{\Delta_{\rm S}\tanh \left( \frac{\Delta_{\rm S}}{2k_B T_{\rm eff}}\right)+\gamma \delta D^{\rm 2D}(B=0,T_{\rm eff})},
    \label{eq:frequency_shift}
\end{align*}
with $\gamma=\gamma/K$.

\begin{figure*}[t!] 	
	\centering
\includegraphics[width=1\textwidth]{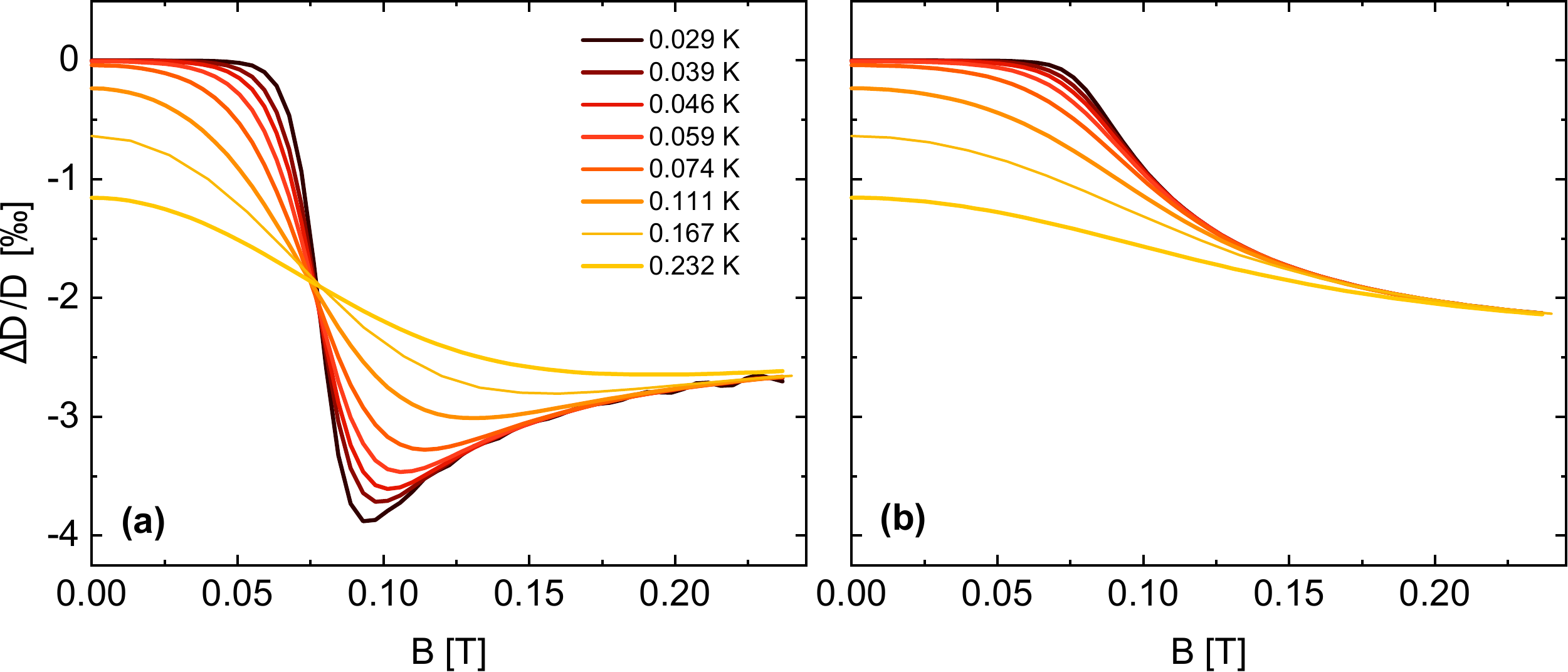} 
    \caption{Magnetic field dependence of the 2DEG+Aluminum stiffness for the model with Zeeman field and SOC, corresponding to various values of $k_BT/\Delta_0$ and $T_{\rm eff}=0.058$~K. Panel \textbf{(a)} refers to the magnetic field perpendicular to the current whereas panel \textbf{(b)} is giving the change in superfluid density for the field being aligned parallel to the inductor. The aluminum contributes with $T=T_{\rm eff}$ and $\Delta_{\rm S}(T_{\rm eff})/\tilde\gamma$=132251 in equation (\ref{t0}).}
    \label{fig:ns_shfit-2}
\end{figure*}

In Fig.~\ref{fig:ns_shfit-1} we show the measured superfluid density $\Delta D/D$ in symbols and the numerical calculated values of the model using equation (\ref{rel-frec}) with lines. This fit in particular corresponds to the scenario of Zeeman+SOC without Doppler shift. From the width of the transition when the applied field is in the perpendicular direction with respect to the wire, we estimated $k_B T_{\text{eff}}/\Delta_0=0.215$. The model establishes only that the transition at zero temperature occurs once the Zeeman energy equals the induced gap. From the magnetic field at which this situation occurs $B^*=0.079$~T, it can be estimated the g-factor $g=2\Delta_0/\mu_B B_c$. Since it is assumed that the induced gap is $\Delta_0=0.08$~meV a g-factor $g\approx35$ can be obtained. A non zero spin-orbit coupling is an essential ingredient in the model without Doppler-shift to obtain an anisotropic response. However, its magnitude only shifts the place in the reciprocal space of the appearance of the Bogoliubov-Fermi surfaces \cite{sano2025thermoelectric}. Hence, we choose to fix the Rashba spin-orbit coupling to the value $\lambda=15$~meV nm. It has been checked that other values do not change the response, nor the BFS's shape or their area. The  only fitting parameter in Eq. (\ref{rel-frec}) is the scaling factor $\gamma$, which is a phenomenological parameter that relates the stiffness of a 2DEG to a magnitude that can be added to the Al stiffness. The specific value is for $\Delta_{\rm S}/\gamma_\parallel=$115992 is fitted from the parallel component of the stiffness, since it is expected that it should be less affected by disorder and therefore, the model should be more reliable for that component. 

Since disorder is present, the simple model for the 2DEG does not reproduce quantitatively the difference in scale of the effect in the parallel and perpendicular direction. Hence, we obtain $\gamma_{\perp}=4 \gamma_\parallel$. The anisotropy is caused by the effect of disorder depending on the presence or absence of Bogoliubov-Fermi surfaces in the direction of the current. 

The model provides additional information on $k_B T/\Delta_0$. In Fig.~\ref{fig:ns_shfit-2}, we show the effect of varying this parameter. The calculations correspond to Eq. (\ref{rel-frec}) with a reference temperature $T_{\text{eff}}=0.058$~K and with the effect of aluminum as a constant shift, i.e. $T=T_{\text{eff}}$ and  $\Delta_{\rm S}(T_{\rm eff})/\gamma_\parallel=$132251. We notice that the transition is sharper if $k_BT/\Delta_0\ll1$ while it gets smoother at $k_BT/\Delta_0\gtrsim0.25$, and even at zero magnetic field only a fraction of the electron density is part of the superfluid. The effect of the temperature is similar in the case of the Doppler scenario. 

\bibliography{Literatur.bib}
\end{document}